\documentclass{aa}
\usepackage{psfig,latexsym,longtable,lscape,rotating}
\begin{document}
\pagenumbering{arabic}
\title{Nearby early--type galaxies with ionized gas.II. Line-strength indices for 18 additional galaxies\footnote{Based on observations obtained at the  European Southern Observatory, 
La Silla, Chile (Programs Nr.~60.A-0647 and 61.A-0406}}
\author{Annibali F.$^1$, Bressan A.$^{1,2}$, Rampazzo R.$^2$, Zeilinger W.W.$^3$}
\institute{
 $^1$ SISSA, via Beirut 4 - 34014 Trieste - Italy \\
 $^2$ INAF - Osservatorio Astronomico di Padova, vicolo dell'Osservatorio 5, 35122 Padova, Italy\\
$^3$Institut f\" ur Astronomie der Universit\" at  Wien, T\" urkenschanzstra$\ss$e 17, A-1180 Wien, Austria 
}
 \offprints{F.~Annibali}
 \mail{annibali@sissa.it}
\date{Received date; accepted date}
\authorrunning{Annibali et al.}
\titlerunning{Nearby early--type galaxies with ionized gas.I Addendum.}
\abstract{Rampazzo et al. \cite{Ramp05} (hereafter Paper I) presented a
data-set of line-strength indices for 50 early-type galaxies
in the nearby Universe. The galaxy sample is biased  toward galaxies showing
emission lines, located in environments corresponding to a broad range
of local galaxy densities, although predominantly in low density
environments. 
The present {\it addendum} to Paper~I enlarges the above data-set 
of line-strength indices by analyzing  18 additional early-type galaxies 
(three galaxies, namely NGC 3607, NGC 5077 and NGC~5898 have been already 
presented in the previous set). 
As in Paper~I, we measured 25 line--strength indices, 
defined by the Lick IDS ``standard'' system 
(Trager et al. \cite{Tra98}; Worthey \& Ottaviani  \cite{OW97}), 
for 7 luminosity weighted apertures 
and  4 gradients of each galaxy. 
This {\it addendum} presents the line-strength data-set and compares it
with the available data in the literature.

 }
\maketitle

\section{Introduction}

The aim of our study is to improve the understanding
of the nature of the ionized gas in early-type galaxies by studying its
physical conditions, the possible ionization mechanisms, 
relations with the other gas components of the Inter Stellar Medium 
(ISM) and the connection
with the stellar population of the host galaxy. 
By investigating issues such as the evolution of stellar
populations and the ISM, we will both explore the complex, 
evolving ecosystem within early--type galaxies 
and build a database of well studied galaxies 
to be used as a reference set for the study of intermediate 
and distant objects. 
Our target is to characterize the stellar populations,
with particular concern for those in the extended emission regions,
through the modelling of the complete (lines and continuum) 
spectrum characteristics. 
This will allow us to constrain the galaxy formation/evolution history. 

The adopted strategy, 
described in Rampazzo et al. \cite{Ramp05} 
(hereafter Paper~I), consists in investigating the galaxy underlying 
stellar populations and the emission line properties
at different galactocentric distances.
The study of
stellar populations in early-type galaxies is of fundamental importance to the
understanding of their evolution with time
(see e.g.  Buzzoni et al. \cite{Buz92}; Worthey \cite{Wor92};  
Gonz\' alez \cite{G93}; Buzzoni et al. \cite{Buz94}; 
Worthey et al. \cite{Wor94}; Leonardi \& Rose \cite {LR96}; 
Worthey \& Ottaviani \cite{OW97}; 
Trager et al. \cite{Tra98}; Longhetti et al. \cite{L98a}; Vazdekis
\cite{Vaz99}; Longhetti et al. \cite{L99}; Longhetti et al. \cite{L00}; 
Trager et al. \cite{Tra00}; Kuntschner \cite{Ku00}; 
Beuing et al.~\cite{Beu02}; 
Kuntschner et al. \cite{Ku02}; Thomas et al. ~\cite{Thom03}; Mehlert et al. 
\cite{Mehl03}) . 

This {\it addendum} intends to enlarge the sample of the early-type galaxies 
analyzed in Paper~I providing the data-set of line--strength indices for 18
objects.
Three galaxies, namely NGC~3607, NGC 5077 and NGC~5898 have
been already presented in the previous set and are used in this
paper in order to check the internal consistency of our results
on repeated observations. Eight galaxies in the present sample, 
namely NGC~3818, NGC~4374, NGC~4697, NGC~5044, NGC~5638, NGC~5812,
NGC~5813 and NGC~5831, belong to the Trager et al. \cite{Tra98}
sample, six of which to the original Gonz\' alez \cite{G93} sample.

This note is organized as follows. Section~2 presents the 
characteristics of the sample of the 18 galaxies. 
Sections~3 and 4 summarize the observations,
the reduction procedure, the line-strength index measurements
at different galactocentric distances and  
the corrections applied to conform the indices
to the Lick IDS system. 
Finally in Section~5 we present the results 
and the comparison with the literature.

\begin{figure*}
\resizebox{8.5cm}{!}
{\psfig{figure=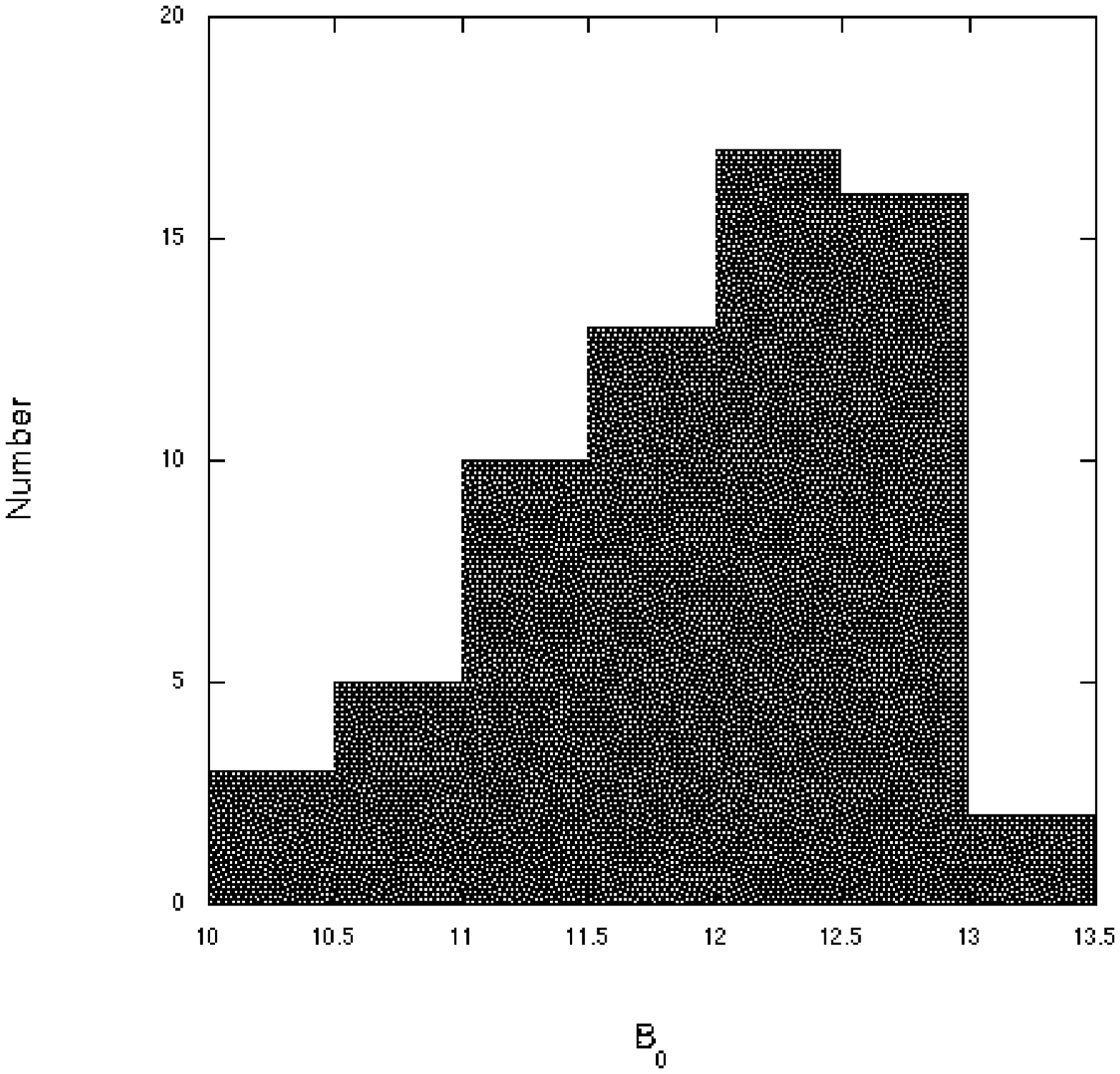,width=16cm,clip=}}
\resizebox{8.5cm}{!}
{\psfig{figure=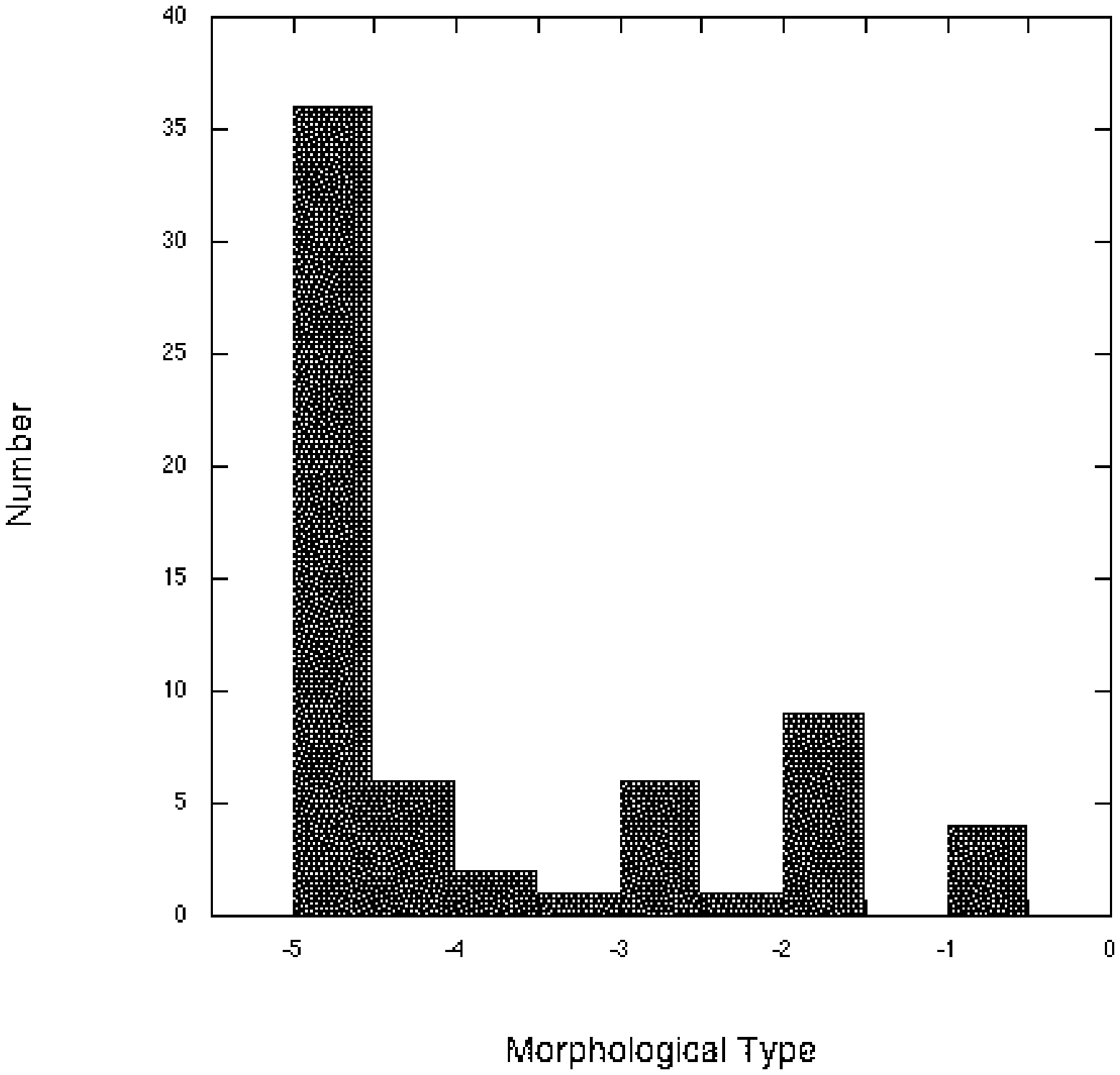,width=16cm,clip=} }
\resizebox{8.5cm}{!}
{ \psfig{figure=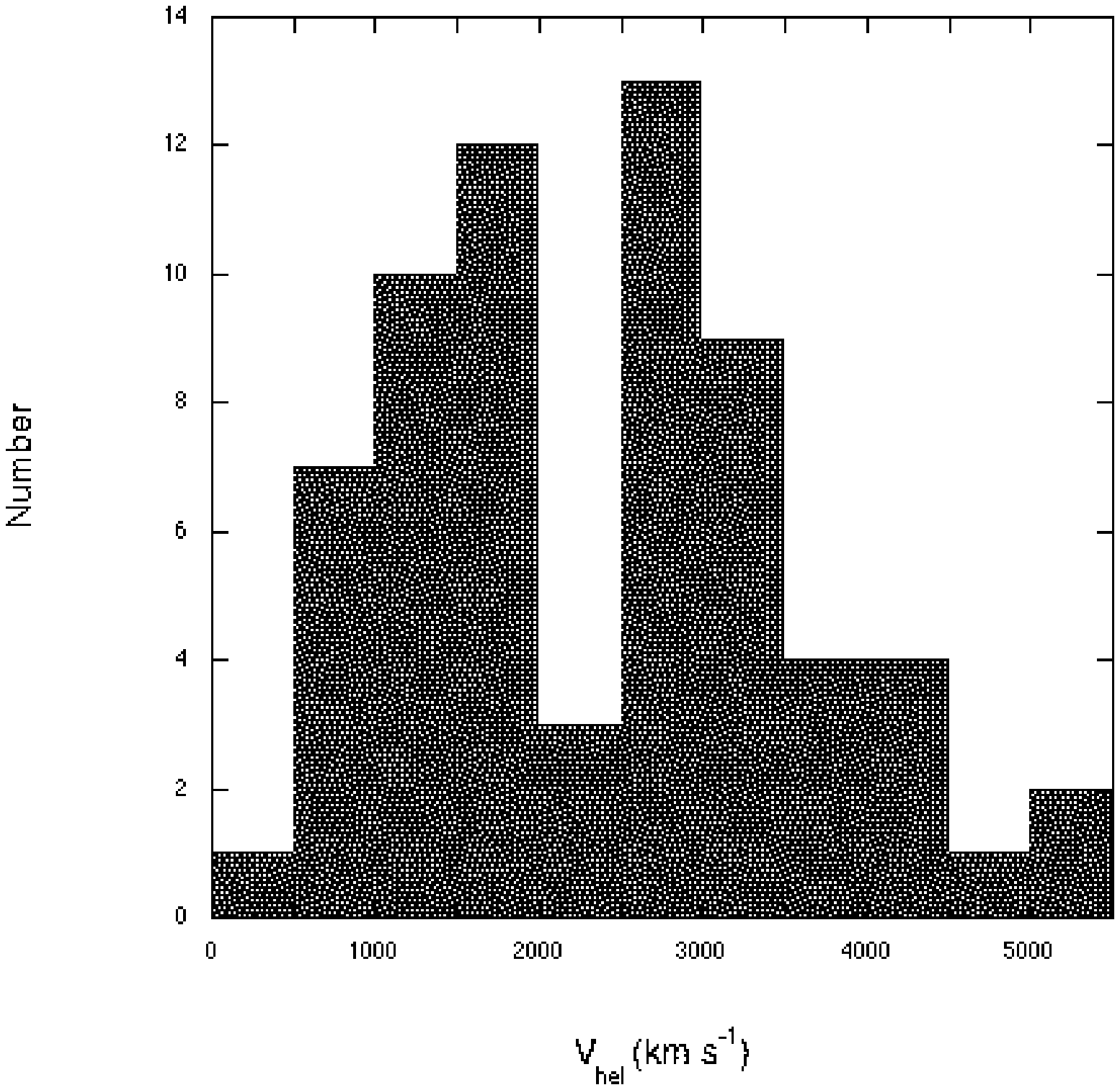,width=16cm,clip=}}
\resizebox{9.1cm}{!}
{\psfig{figure=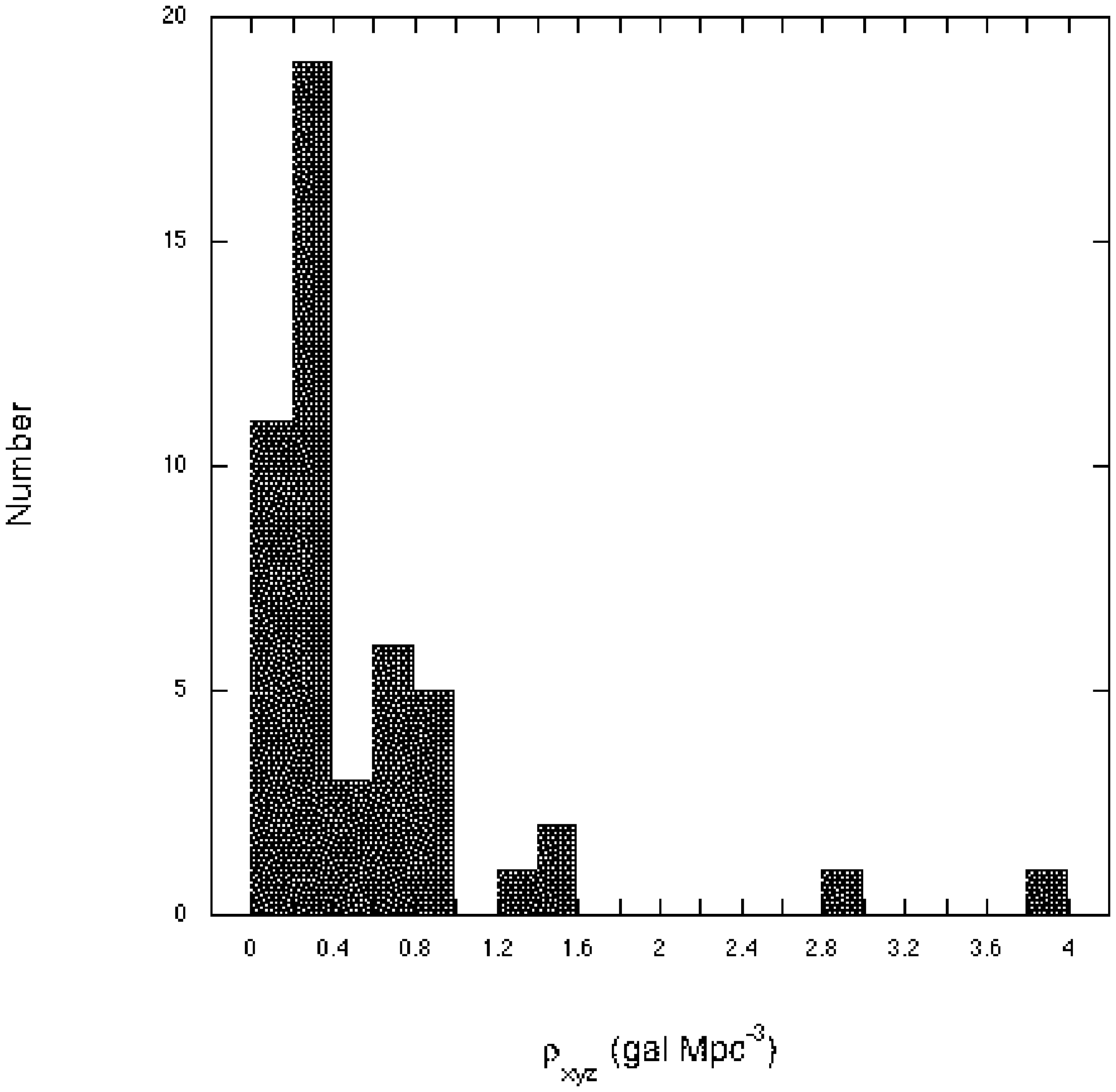,width=16cm,clip=} } 
\caption{Distribution of 
B-magnitudes (top left panel), morphological types (top right panel), 
heliocentric velocity (bottom left panel) and galaxy density 
(bottom right panel) for the {\it total sample}  (50 galaxies of Paper~I 
+ 15 new galaxies of this paper).} 
\label{fig1}
\end{figure*}


\begin{figure}
\psfig{figure=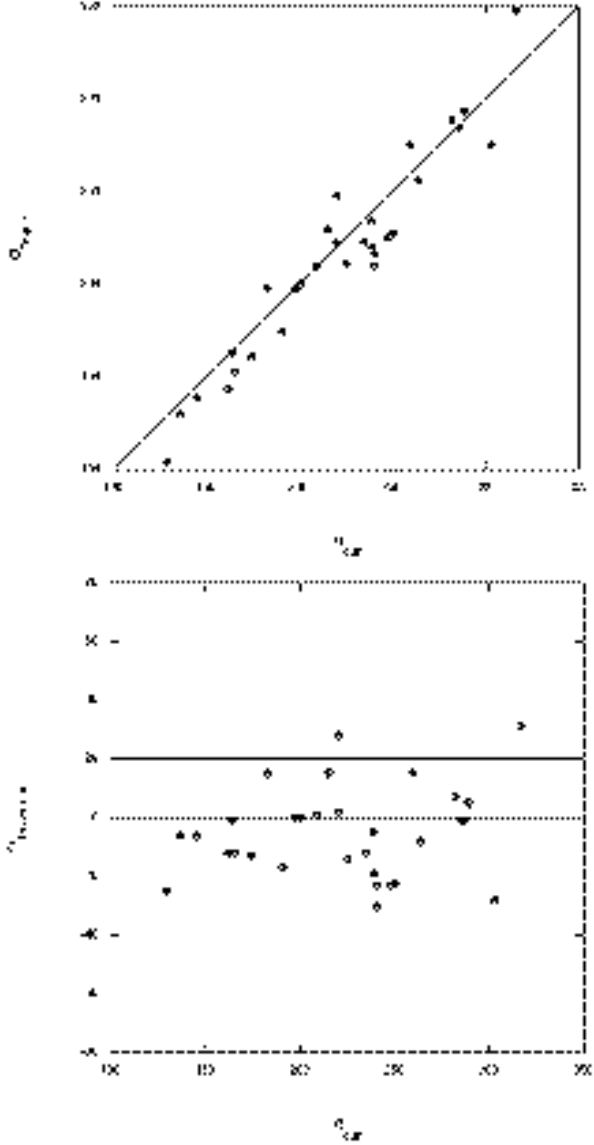,width=9cm,clip=}
\caption{Comparison between central velocity dispersions 
used in this paper and in Trager et al. (\cite{Tra98}). The lines indicate an 
average error (20 km~s$^{-1}$) in the central velocity dispersion measurements.} 
\label{fig2}
\end{figure}

\begin{table*}
\begin{tabular}{lcrllrrcrrclc}
& & & & & & & & & &  & &\\
\multicolumn{13}{c}{\bf Table 1} Overview of the observed sample \\
\hline\hline
\multicolumn{1}{c}{ident}
& \multicolumn{2}{c}{R.A. (2000) Dec.}
& \multicolumn{1}{c}{RSA}
& \multicolumn{1}{c}{RC3}
& \multicolumn{1}{c}{P.A.}
& \multicolumn{1}{c}{B$_0$}
& \multicolumn{1}{c}{(B-V)$_0$}
 & \multicolumn{1}{c}{(U-B)$_0$}
& \multicolumn{1}{c}{V$_{hel}$}
& \multicolumn{1}{c}{r$_e$}
& \multicolumn{1}{c}{$\rho_{xyz}$} 
& \multicolumn{1}{c}{$\epsilon$} \\
 & & & & & & & & & &  & &\\
 \hline\hline
 & & & & & & & & & &  & &\\
 NGC~3607 & 11 16 54.3 &  18 03 10 & S03(3)     & SA(s)0:       & 120 & 11.08 & 0.88 & 0.43 &  934 &  43.4  & 0.34 & 0.11 \\
 NGC~3818 & 11 41 57.3 &-06 09 20 &    E5        & E5              & 103 & 12.47 & 0.91 &         &1701 &  22.2  & 0.20 & 0.36 \\
 NGC~4374 & 12 25 03.7 &  12 53 13 &    E1        & E1              & 135 & 10.01 & 0.94 & 0.50 &1060 &  50.9  & 3.99 & 0.11\\
 NGC~4696 & 12 48 49.3 &-41 18 40 &    (E3)       & E+1 pec     &   95 & 10.85 &         &         &2958 &   85.0 & 0.00 & 0.34\\
 NGC~4697 & 12 48 35.9 &-05 48 03 &     E6       & E6              &   70 & 10.07 & 0.89 & 0.39 &1241 &  72.0  & 0.60 & 0.32\\
 NGC~5011 & 13 12 51.8 &-43 05 46 &     E2      & E1-2          & 154 & 11.90 & 0.89 &         &3104 &   23.8 & 0.27 & 0.15\\
 NGC~5044 & 13 15 24.0 &-16 23 08 &     E0      & E0              &   11 & 11.67 &         &         &2704 &  82.3  & 0.38 & 0.11\\
 NGC~5077 & 13 19 31.6 &-12 39 26 & S01/2(4)  & E3+           &  11 & 12.52 & 0.98 & 0.54 & 2764 & 22.8   & 0.23 & 0.15\\
 NGC~5090 & 13 21 12.8 &-43 42 16 &     E2      & E2              & 136& 11.97 &         &         & 3421 & 62.4   &         & 0.15\\
 NGC~5193 & 13 31 53.5 &-33 14 04 &    S01(0)  & E pec          &  71 & 12.35 & 0.86 &         & 3711 & 26.7   &         & 0.07\\
 & & & & & & & & & &  & &\\ 
 NGC~5266 & 13 43 02.1 &-48 10 10 &S03(5) pec&SA0-:          & 109& 11.50 &         &         & 3074 & 76.7   & 0.35 & 0.31\\
 NGC~5638 & 14 29 40.4 &  03 14 00 &   E1       &E1               &150 & 12.06 & 0.91 & 0.46 & 1676 & 28.0  &  0.79 & 0.11\\
 NGC~5812 & 15 00 55.7 &-07 27 26 &    E0        & E0              &130 & 11.83 & 0.94 &         & 1930 & 25.5  &  0.19 & 0.13\\
 NGC~5813 & 15 01 11.2 & 01 42 07  &   E1        &  E1-2         &145 & 11.42 & 0.94 & 0.52 & 1972 & 57.2  &  0.88 & 0.28\\
 NGC~5831 & 15 04 07.0 & 01 13 12  &   E4        & E3              &  55 & 12.31 & 0.92 & 0.55 & 1656 & 25.5  &  0.83 & 0.13\\
 NGC~5898 & 15 18 13.6 & -24 05 51& S02/3(0)  & E0             &  30  & 12.41 & 0.95 &         & 2267 & 22.2  & 0.23  & 0.07\\
 NGC~6758 & 19 13 52.3 & -56 18 36& E2 (merger) &  E+:           &  121& 12.31 & 0.96 & 0.44 & 3404 & 20.3  &          & 0.22\\
 NGC~6776 & 19 25 19.1 &-63 51 37 & E1 pec &  E+ pec      &   15 &  12.71& 0.86& 0.44  & 5480 & 17.7  &         &  0.17\\
& & & & & & & & & &  & &\\
\hline\hline
\end{tabular}
\label{table1}

\medskip
{{\bf Notes}: The value of r$_e$ of NGC~4696, NGC 5090 and NGC 5266 have been derived from ESO-LV (Lauberts \& Valentijn (\cite{LV89})
while the value for NGC 5044 is derived from Goudfrooij (\cite{Gou94}) .}
\end{table*}

\section{Characterization of the sample}

As the original sample of 50 early--type galaxies presented in Paper~I, 
this sample of 18 galaxies is selected
from a compilation of galaxies showing ISM traces in at least one of the
following bands: IRAS 100 $\mu$m, X-ray, radio, HI and CO (Roberts et
al. \cite{Ro91}).  All galaxies belong to {\it Revised Shapley Ames
Catalog of Bright Galaxies (RSA)} (Sandage \& Tammann \cite{RSA}) and
have a redshift of less than 5500 km~s$^{-1}$. The sample should then be
biased towards objects that might be expected to have ongoing and recent
star formation, at least in small amounts, because of the presence of emission
lines. The emission should come from a combination of active galactic 
nuclei and star formation regions within the galaxies.
Table~1 summarizes  the basic characteristics of the
galaxies available from the literature.  Column (1)
provides the identification; columns (2) and (3) the R.A. \& Dec.
coordinates; columns (4) and (5) the galaxy morphological classifications 
according to the RSA (Sandage \& Tamman \cite{RSA}) and RC3 (de Vaucouleurs et
al. \cite{RC3}) respectively. 
Columns (6), (7), (8) and (9) give the position angle of
the isophotes along the major axis, the total corrected magnitude and the total
(B-V) and (U-B) corrected colors from RC3 respectively. The
heliocentric systemic velocity from HYPERCAT
({\tt http://www-obs.univ-lyon1.fr/hypercat}) is reported in column
(10). The effective radius, derived from A$_e$,  the diameter of the
effective aperture from RC3, is given in column (11). A measure of
the richness of the environment, $\rho_{xyz}$ (galaxies~Mpc$^{-3}$), 
surrounding each galaxy is reported in column (12) (Tully
\cite{Tu88}). Column (13) lists the average ellipticity of the
galaxy as obtained from HYPERCAT.  

Figure~\ref{fig1} summarizes the basic characteristics of the {\it total
sample} (50 + 15 new galaxies of this paper) and in particular, in
the right bottom panel, provides evidence that a large fraction of galaxies
is in low density environments. 

Excluding NGC~3607, NGC~5077 and NGC~5898 already discussed in Paper~I, 
seven of the new  galaxies in the present sample belong to the sample of
73 luminous early--type galaxies selected from the RC3 catalogue by
Macchetto et al. (\cite{Mac96}) for a characterization of the extended
emission region in H$\alpha$ + [NII] emission lines. 
The seven galaxies are NGC~5044, NGC~5090, NGC~5812, BGC~5813, 
NGC~5831, NGC~6758 and NGC~6776. 
Macchetto et al. (\cite{Mac96}) showed
that in three of these galaxies, namely NGC~5044, NGC~5813 and 
NGC~6776, the shape of the emission region is filamentary (F). In the
remaining galaxies the ionized gas morphology is more regular,
(NGC~6758), or reminiscent of a small disk (SD), (NGC~5090, NGC~5812 and
NGC~5831).

In Appendix A we have collected  individual notes on galaxies,
emphasizing kinematical studies of the ionized gas component, its
correlation with the stellar body and its possible origin. Information
for NGC~3607, NGC~5077 and NGC~5898 are given in Paper~I. 

Several classes of galaxies are present in this additional sample: 
interacting or post-interacting galaxies, galaxies showing evidence of 
kinematical decoupling between galaxy sub-components, elliptical galaxies with 
a dust lane along the minor axis, radio galaxies and galaxies hosting an AGN nucleus. 
To summarize the individual notes in Appendix A,  the sample contains one
galaxy showing a shell structure, namely NGC~6776, and a galaxy, NGC~5266,
considered an "old merger" of two spirals having a large HI content.  
Six galaxies (namely NGC~4697, NGC~5077, NGC~5266, NGC~5813, NGC~5898 and NGC~6758)
have a peculiar kinematical behaviour, e.g. counterrotation of stars vs. 
gas and/or stars vs. stars. Three galaxies (namely NGC~4374, NGC~4696 and NGC~5090) 
are FRI type radio sources. Often the ionized gas is associated to dust-lane complexes 
(e.g. NGC~4374, NGC~4696, NGC~5266 and NGC~5813,  
see e.g. Goudfrooij \cite{Gou94}; Goudfrooij \cite{Gou98}). 
 Some objects, as NGC~5638 and NGC~5813, seem finally "unperturbed" elliptical. 

\section{Observations and data reduction}

\subsection{Observations}

Galaxies were observed between May 10 and 13th, 1999 at the 1.5m ESO telescope 
(La Silla), equipped with a Boller \& Chivens spectrograph and a UV coated 
CCD Fa2048L 
(2048$\times$2048) camera (ESO CCD \#39). Details of the observations and
typical seeing during the run are reported in  Table~2. Table~3
provides a journal of observations, i.e. the object identification (column 1),
the slit position angle oriented North through
East (column 2) and the total exposure time (column 3) and observing 
conditions (column 4). The spectroscopic
slit was oriented along the galaxy major axis for most  observations. HeAr-FeNe
calibration frames were taken before and after each exposure to allow an
accurate wavelength calibration. 

\subsection{Data reduction, extraction of apertures and gradients} 

We adopted the same data reduction procedure 
described in Paper~I. 
We recall here that the fringing seriously affected observations 
longward of $\approx$ 7300\AA. 
After accurate flat-fielding correction, we considered the wavelength 
range 3700 - 7250 \AA  \ for further use. Multiple spectra for a given galaxy 
were co-added and flux calibrated using a sequence of spectrophotometric 
standard stars.

The definition of the apertures and gradients and their extraction 
procedure  are detailed in Paper~I. 
Summarizing, we have extracted flux-calibrated spectra 
along the slit in seven circular concentric  regions, hereafter "apertures", 
and in four adjacent regions, hereafter "gradients". The seven 
luminosity weighted apertures, corrected for the galaxy ellipticity, 
have radii of
1.5\arcsec, 2.5\arcsec, 10\arcsec,  r$_e$/10, r$_e$/8, r$_e$/4 and r$_e$/2.
The four gradients are derived
in the regions 0 $\leq$ r $\leq$r$_e$/16, r$_{e}$/16 $\leq$ r $\leq$r$_e$/8, 
r$_{e}$/8 $\leq$ r $\leq$r$_e$/4 and r$_{e}$/4 $\leq$ r $\leq$r$_e$/2.

\begin{table*}
\small{
\begin{tabular}{ll}
& \\
\multicolumn{2}{c}{\bf Table 2. } Observing parameters \\
\hline\hline

\hline
Date of Observations & 10-13 May 99 \\
Observer  & Zeilinger W.\\
Spectrograph & B \& C grating \#25\\
Detector & ESO CCD \#39, Loral 2K UV flooded\\
Pixel size ($\mu$m) & 15  \\
Scale along the slit (\arcsec/px$^{-1}$) & 0.82 \\
Slit length (\arcmin) & 4.5 \\
Slit width (\arcsec) & 2  \\
Dispersion(\AA\ mm$^{-1}$) & 187 \\
Spectral Resolution (FWHM at 5200 \AA\ ) (\AA ) & 7.6  \\
Spectral Range (\AA ) & 3622-9325 \\
Seeing Range(FWHM) (\arcsec) & 1.2-2  \\
Standard stars & Feige 56, ltt 1788, ltt 377 \\
\hline
\end{tabular}}
\label{table2}
\end{table*}


\begin{table*}
\small{
\begin{tabular}{lccl}
& & & \\
\multicolumn{4}{c}{\bf Table 3. } Journal of galaxy observations \\
\hline\hline
\multicolumn{1}{c}{ident.}&
\multicolumn{1}{c}{Slit PA} &
\multicolumn{1}{c}{t$_{exp}$}   &
\multicolumn{1}{c}{observational notes}  \\
\multicolumn{1}{c}{}& 
\multicolumn{1}{c}{[deg]} & 
\multicolumn{1}{c}{[sec]}  &
\multicolumn{1}{c}{}  \\
\hline
NGC~3607  &   120  & 2$\times$1800 &\\
NGC~3818  &   103  & 2$\times$1800 &\\
NGC~4374  &   135  & 2$\times$1800 & clouds\\
NGC~4696  &   95  & 2$\times$1800 &  clouds\\
NGC~4697 &   70  & 2$\times$1800  &   \\
NGC~5011&   154& 2$\times$1800 & clouds\\
NGC~5044&   11  & 2$\times$1800 & low S/N, clouds\\
NGC~5077 &   11 & 2$\times$1800 & low S/N \\
NGC~5090 &   136  & 2$\times$1800 & clouds, spectrum includes NGC 5091 spiral\\
NGC~5193&   71  & 2$\times$1800  & clouds, spectrum includes a spiral galaxy in the field \\
 & & \\
NGC~5266&   109  & 2$\times$1800 & low S/N\\
NGC~5638&   150  & 2$\times$1800 &\\
NGC~5812&   130  & 2$\times$1800 & clouds\\
NGC~5813&   145  & 2$\times$1800 & \\
NGC~5831&   55  & 2$\times$1800   &\\
NGC~5898&   30  & 2$\times$1800   & clouds, strong wind\\
NGC~6758&   121  & 2$\times$1800 & \\
NGC~6776&   15  & 2$\times$1800   & \\
\hline
\end{tabular}}
\label{table3}
\end{table*}

\begin{table*}
\small{
\begin{tabular}{lcccllcccl}
& & & & & & &  \\
\multicolumn{8}{c}{\bf Table 4.} Velocity dispersion values adopted in the correction of the line--strength indices \\
\hline\hline
\multicolumn{1}{l}{Ident.} &
\multicolumn{1}{c}{$\sigma_{r_e/8}$} &
\multicolumn{1}{c}{$\sigma_{r_e/4}$} &
\multicolumn{1}{c}{$\sigma_{r_e/2}$} &
\multicolumn{1}{l}{Ref.} &
\multicolumn{1}{l}{Ident.} &
\multicolumn{1}{c}{$\sigma_{r_e/8}$} &
\multicolumn{1}{c}{$\sigma_{r_e/4}$} &
\multicolumn{1}{c}{$\sigma_{r_e/2}$} &
\multicolumn{1}{l}{Ref.} \\
\multicolumn{1}{c}{} &
\multicolumn{1}{c}{[km~s$^{-1}$]} &
\multicolumn{1}{c}{[km~s$^{-1}$]} &
\multicolumn{1}{c}{[km~s$^{-1}$]} &
\multicolumn{1}{c}{} &
\multicolumn{1}{c}{} &
\multicolumn{1}{c}{[km~s$^{-1}$]} &
\multicolumn{1}{c}{[km~s$^{-1}$]} &
\multicolumn{1}{c}{[km~s$^{-1}$]} &
\multicolumn{1}{c}{}\\
\hline
& & & & & & &  & &  \\
NGC~3607   & 220  & 210  & 195  & CMP00     & NGC~5266  & 199 & 140  &         & CDB93 \\
NGC~3818   & 191  & 165  & 138  &   SP97      & NGC~5638  &  165&  151 &  132 & Dav83 \\
NGC~4374   & 282  & 283  & 236  &  Dav81     & NGC~5812  & 200 & 204 &   172 & B94  \\
NGC~4696   & 254  & 190  &         &  CDB93    & NGC~5813  & 239 & 220 &  220 & CMP00  \\
NGC~4697   & 174  & 148  & 130  & Pink03     & NGC~5831  & 164 & 143 &  124 & Dav83  \\
NGC~5011   & 249  & 220  & 200  & CDB93     & NGC~5898  & 220 & 183 &  172 & CMP00  \\
NGC~5044   & 239  & 220  &         & CMP00     & NGC~6758  & 242 & 210 &  205 &  CMP00        \\
NGC~5077   & 260  & 239  & 228  &CMP00      & NGC~6776  & 242 & 214 &  207 &  L98 \\
NGC~5090   & 269  & 250  &         & CDB93     &                    &        &        &         &   \\
NGC~5193   & 209  &         &         &                 &                   &        &         &        &          \\
 & & & & & & & & &    \\
\hline
\end{tabular}}
\label{table4}

\medskip
{\bf Notes}:  the average central value, obtained from the on-line compilation
HYPERCAT (http://www-obs.univ-lyon1.fr/hypercat/), is adopted for
$\sigma_{r_e/8}$. At larger radii,  $\sigma_{r_e/4}$ and
$\sigma_{r_e/2}$ are obtained from the references
quoted in columns 5 and 10. Legend: 
B94 = Bertin et al.  (\cite{B94}); 
CDB93 = Carollo et al. (\cite{CDB93}); 
CMP00 = Caon et al. (\cite{CMP00}); 
Dav81 = Davies (\cite{Davies81});
Dav83 = Davies et al. (\cite{Davies83});
L98 = Longhetti et al. (\cite{L98b}); 
Pink03 =  Pinkney et al. (\cite{Pink03}).
SP97 = Simien \& Prugniel (\cite{SP97})
\end{table*}


\begin{table}
\small{
\begin{tabular}{lcccc}
& &  & &\\
\multicolumn{5}{c}{\bf Table 5.} H$\beta$ corrections for apertures
\\
\hline\hline
\multicolumn{1}{c}{Galaxy}
&\multicolumn{1}{c}{aperture}
& \multicolumn{1}{c}{EWO[III]}
& \multicolumn{1}{c}{Quality}
& \multicolumn{1}{c}{EWH$\alpha$}
\\
 \hline 
    NGC~3607 & 0 &  -0.342  &1.000  & -0.651  \\
    NGC~3607 & 1 &  -0.291 & 1.000    &-0.582\\
    NGC~3607 & 2 &  -0.255 & 1.000    &-0.505\\  
    NGC~3607 & 3 &  -0.246 &1.000   & -0.519  \\  
    NGC~3607 & 4 &  -0.217 & 1.000  & -0.529\\ 
    NGC~3607 & 5 &  -0.139 &0.000   & -0.457\\  
    NGC~3607 & 6 &  -0.104 &0.000&   -0.271\\ 
    NGC~3818 & 0 &  -0.173 &0.000 &  -0.141  \\ 
    NGC~3818 & 1 &  -0.154& 0.000 &  -0.142\\
    NGC~3818 & 2 & -0.182 & 0.000 &   -0.154 \\ 
    NGC~3818 & 3 & -0.182 & 0.000 &   -0.162 \\  
    NGC~3818 & 4 & -0.227 & 1.000 & -0.222 \\ 
    NGC~3818 & 5 & -0.178 & 0.000 &   -0.235\\  
    NGC~3818 & 6 & -0.140 &0.000 &   -0.295 \\  
    ........       & .....&   ......   &  ......     &  .......  \\
\hline
\end{tabular}}
\label{table5}

\medskip
The table provides the object identification (column 1), the 
number of the aperture (column 2),
the corresponding radius of which 
is given in Table~9, and the derived EW of the O[III] and H$\alpha$ 
emissions (columns 3 and 5 respectively).
Column 4 gives the quality of the measure, obtained as the ratio between the
estimated emission (F$_{gal}$ - F$_{templ}$) and the variance
of the spectrum in the O[III] wavelength region, $\sigma_\lambda$. 
When the emission is lower than the variance, the quality is set to 0. 
When the emission is between 1 and 2  $\sigma_\lambda$ or larger 
than 2  $\sigma_\lambda$, the quality is set to 1 and 2 respectively. 
The entire table is given is electronic form.
\end{table}

\begin{table}
\small{
\begin{tabular}{lcccc}
& &  & &\\
\multicolumn{5}{c}{\bf Table 6.} H$\beta$ corrections for gradients
\\
\hline\hline
\multicolumn{1}{c}{Galaxy}
&\multicolumn{1}{c}{aperture}
& \multicolumn{1}{c}{EWO[III]}
& \multicolumn{1}{c}{Quality}
& \multicolumn{1}{c}{EWH$\alpha$}
\\
 \hline         
    NGC~3607 & 0 &   -0.369  &    1.000  &   -0.608 \\  
    NGC~3607  &1 &   -0.183  &    0.000  &   -0.445\\  
    NGC~3607 & 2 &   -0.263  &    1.000  &   -0.434\\  
    NGC~3607  & 3 &   0.117  &    0.000  &   -0.129\\
    NGC~3818  & 0 &  -0.1947 &    0.000  &   -0.135\\
    NGC~3818  & 1 &  -0.220  &     1.000 &   -0.153\\  
    NGC~3818  & 2 &  -0.352  &     1.000 &   -0.251\\  
    NGC~3818  & 3 &  -0.211  &     0.000 &   -0.428\\ 
     ....           & ... &   .....     &   .....     &  .....  \\
\hline
\end{tabular}}
\label{table6}

\medskip
The table provides the object identification (column 1), the 
number of the gradient (column 2) 
(0=0 $\leq$ r $\leq$r$_e$/16, 1 = r$_{e}$/16 $\leq$ r $\leq$r$_e$/8, 
2 = r$_{e}$/8 $\leq$  r $\leq$r$_e$/4 and 
3 = r$_{e}$/4 $\leq$ r $\leq$r$_e$/2), 
and the derived EW of the O[III] and H$\alpha$ emissions (columns 3 and 5
respectively). 
Column 4 gives the quality of the measure, obtained as the ratio between the
estimated emission (F$_{gal}$ - F$_{templ}$) and the variance
of the spectrum in the O[III] wavelength region, $\sigma_\lambda$. 
When the emission is lower than the variance, the quality is set to 0. 
When the emission is between 1 and 2  $\sigma_\lambda$ or larger than 
2  $\sigma_\lambda$, the quality is set to 1 and 2 respectively. 
The entire table is given in electronic form.
\end{table}


\begin{table}
\small{
\begin{tabular}{lcccc}
& &  & & \\
\multicolumn{5}{c}{\bf Table 7}  $\alpha$ and $\beta$ coeff. for transformation into Lick system \\
\hline\hline
\multicolumn{1}{c}{Index}
& \multicolumn{1}{c}{$\alpha$}
& \multicolumn{1}{c}{$\beta$}
& \multicolumn{1}{c}{aver. disp.}
& \multicolumn{1}{c}{unit}
 \\
 & &  & &\\
\hline
CN$_1$           &         1.059  &         0.023  & 0.025   & mag\\
CN$_2$           &         1.035  &         0.030  & 0.023   & mag \\
Ca4227        &         1.317  &         0.408  & 0.396   & \AA \\
G4300         &         1.105  &         0.179 &  0.310  & \AA\\
Fe4383        &         0.963  &         1.169  & 0.772   & \AA \\
ca4455        &         0.451  &         1.844  & 0.341   & \AA \\
Fe4531        &         1.289  &        -0.299  &0.437    & \AA\\
Fe4668        &         0.976  &         0.128  &0.653    & \AA \\
H$\beta$    &         1.064  &        -0.196   &0.166    & \AA \\
Fe5015        &         1.031  &         0.804  & 0.396   & \AA \\
Mg$_1$           &         1.014  &         0.015  &  0.009  & mag \\
Mg$_2$           &         0.998  &         0.020   & 0.012  &  mag \\
Mgb           &         1.014  &         0.417  &  0.241  & \AA \\
Fe5270        &         1.058  &         0.270  & 0.240   & \AA\\
Fe5335        &         0.990  &         0.356  & 0.249   & \AA \\
Fe5406        &         1.005  &         0.282  & 0.151   & \AA \\
Fe5709        &         1.321  &        -0.270  & 0.174   & \AA \\
Fe5782        &         1.167  &        -0.075 &  0.165  & \AA \\
NaD           &          1.003  &         0.027  & 0.245   & \AA \\
TiO$_1$          &          0.997  &         0.004  & 0.006   &  mag\\
TiO$_2$          &          1.003  &        -0.001   &0.008    &   mag \\
H$\delta_{A}$ &         1.136  &        -0.622   &1.087    & \AA \\
H$\gamma_{A}$ &        0.990  &         0.518   & 0.734   & \AA \\
H$\delta_{F}$ &         1.059  &        -0.036   & 0.503   & \AA \\
H$\gamma_{F}$ &         1.011  &         0.458   & 0.745   & \AA \\
\hline
\end{tabular}}
\label{table8}

\medskip
The table provides the coefficients $\alpha$ and $\beta$ of the transformation
into the Lick system ($EW_{Lick} = \beta + \alpha \times EW_{our}$)
derived in Paper~I.
\end{table}

\begin{table*}
\tiny{
\begin{tabular}{lrrrrrrrrrrrrrrrr}
 & & & & & & & & & & & & & & &   \\
\multicolumn{16}{c}{\bf Table 8 } Fully corrected line--strength
indices for the apertures  \\
\hline\hline
\multicolumn{1}{c}{galaxy} &
\multicolumn{1}{c}{iz} &
\multicolumn{1}{c}{ie} &
\multicolumn{1}{c}{r$_1$} &
\multicolumn{1}{c}{r$_2$} &
\multicolumn{1}{c}{CN$_1$} &
\multicolumn{1}{c}{CN$_2$} &
\multicolumn{1}{c}{Ca4227} &
\multicolumn{1}{c}{G4300} &
\multicolumn{1}{c}{Fe4383} &
\multicolumn{1}{c}{Ca4455} &
\multicolumn{1}{c}{Fe4531} &
\multicolumn{1}{c}{Fe4668} &
\multicolumn{1}{c}{H$\beta$} &
\multicolumn{1}{c}{Fe5015}&
\multicolumn{1}{c}{other}  \\
\multicolumn{1}{c}{galaxy} &
\multicolumn{1}{c}{iz} &
\multicolumn{1}{c}{ie} &
\multicolumn{1}{c}{r$_l$} &
\multicolumn{1}{c}{r$_e$} &
\multicolumn{1}{c}{eCN$_1$} &
\multicolumn{1}{c}{eCN$_2$} &
\multicolumn{1}{c}{eCa4227} &
\multicolumn{1}{c}{eG4300} &
\multicolumn{1}{c}{eFe4383} &
\multicolumn{1}{c}{eCa4455} &
\multicolumn{1}{c}{eFe4531} &
\multicolumn{1}{c}{eFe4668} &
\multicolumn{1}{c}{eH$\beta$}&
\multicolumn{1}{c}{eFe5015}&
\multicolumn{1}{c}{indices} \\
\hline
   NGC3607 &  0 &   0 &      0.000 &      0.035 &      0.146 &      0.181 &      1.630 &      5.940 &      6.040 &      2.420 &      3.610 &      8.580 &      1.480 &      5.720  & ...\\ 
   NGC3607 &  0 &   1 &      0.023 &     43.400 &      0.004 &      0.005 &      0.082 &      0.139 &      0.165 &      0.085 &      0.135 &      0.211 &      0.088 &      0.212 & ... \\
   NGC3607 &  1 &   0 &      0.000 &      0.058 &      0.139 &      0.172 &      1.570 &      5.900 &      5.870 &      2.320 &      3.810 &      8.260 &      1.490 &      5.680 & ... \\ 
   NGC3607 &  1 &   1 &      0.037 &     43.400 &      0.003 &      0.003 &      0.055 &      0.093 &      0.152 &      0.079 &      0.138 &      0.225 &      0.089 &      0.231 & ... \\ 
   NGC3607 &  2 &   0 &      0.000 &      0.100 &      0.124 &      0.153 &      1.600 &      5.890 &      5.830 &      2.310 &      3.900 &      7.920 &      1.500 &      5.730 & ... \\ 
   NGC3607 &  2 &   1 &      0.060 &     43.400 &      0.003 &      0.003 &      0.055 &      0.095 &      0.154 &      0.079 &      0.138 &      0.228 &      0.090 &      0.233 & ... \\ 
   NGC3607 &  3 &   0 &      0.000 &      0.115 &      0.119 &      0.150 &      1.580 &      5.830 &      5.800 &      2.310 &      3.860 &      7.750 &      1.500 &      5.760 & ... \\ 
   NGC3607 &  3 &   1 &      0.068 &     43.400 &      0.003 &      0.003 &      0.056 &      0.096 &      0.155 &      0.080 &      0.139 &      0.230 &      0.090 &      0.235 & ...  \\ 
   NGC3607 &  4 &   0 &      0.000 &      0.125 &      0.116 &      0.148 &      1.560 &      5.820 &      5.810 &      2.320 &      3.860 &      7.670 &      1.480 &      5.740 & ... \\ 
   NGC3607 &  4 &   1 &      0.073 &     43.400 &      0.003 &      0.003 &      0.056 &      0.097 &      0.155 &      0.080 &      0.139 &      0.231 &      0.091 &      0.236 & ... \\ 
   NGC3607 &  5 &   0 &      0.000 &      0.250 &      0.097 &      0.129 &      1.500 &      5.770 &      5.680 &      2.330 &      3.710 &      6.970 &      1.400 &      5.600 & ... \\ 
   NGC3607 &  5 &   1 &      0.127 &     43.400 &      0.003 &      0.004 &      0.062 &      0.106 &      0.170 &      0.087 &      0.153 &      0.257 &      0.100 &      0.259 & ... \\ 
   NGC3607 &  6 &   0 &      0.000 &      0.500 &      0.091 &      0.120 &      1.640 &      5.750 &      5.790 &      2.380 &      3.560 &      6.450 &      1.440 &      5.570 & ... \\ 
   NGC3607 &  6 &   1 &      0.226 &     43.400 &      0.006 &      0.007 &      0.106 &      0.170 &      0.214 &      0.118 &      0.198 &      0.287 &      0.127 &      0.272 & ... \\ 
   NGC3818 &  0 &   0 &      0.000 &      0.068 &      0.143 &      0.183 &      1.650 &      6.170 &      5.800 &      2.320 &      3.620 &      7.400 &      1.570 &      5.800 & ... \\ 
   NGC3818 &  0 &   1 &      0.042 &     22.200 &      0.004 &      0.005 &      0.074 &      0.127 &      0.203 &      0.113 &      0.189 &      0.268 &      0.114 &      0.289 & ... \\ 
   NGC3818 &  1 &   0 &      0.000 &      0.100 &      0.130 &      0.162 &      1.660 &      6.090 &      5.760 &      2.300 &      3.690 &      7.100 &      1.610 &      5.800 & ... \\ 
   NGC3818 &  1 &   1 &      0.059 &     22.200 &      0.004 &      0.005 &      0.072 &      0.125 &      0.199 &      0.112 &      0.186 &      0.264 &      0.111 &      0.281 & ... \\ 
   NGC3818 &  2 &   0 &      0.000 &      0.113 &      0.128 &      0.158 &      1.650 &      6.050 &      5.730 &      2.300 &      3.670 &      6.970 &      1.660 &      5.830 & ... \\ 
   NGC3818 &  2 &   1 &      0.064 &     22.200 &      0.003 &      0.003 &      0.058 &      0.107 &      0.164 &      0.074 &      0.134 &      0.217 &      0.094 &      0.185 & ... \\ 
   NGC3818 &  3 &   0 &      0.000 &      0.125 &      0.125 &      0.156 &      1.690 &      6.030 &      5.700 &      2.290 &      3.630 &      6.850 &      1.680 &      5.810 & ... \\ 
   NGC3818 &  3 &   1 &      0.070 &     22.200 &      0.003 &      0.004 &      0.058 &      0.106 &      0.164 &      0.074 &      0.134 &      0.217 &      0.094 &      0.186 & ... \\ 
   NGC3818 &  4 &   0 &      0.000 &      0.225 &      0.100 &      0.130 &      1.570 &      5.960 &      5.320 &      2.270 &      3.390 &      5.990 &      1.760 &      5.560 & ... \\
   NGC3818 &  4 &   1 &      0.107 &     22.200 &      0.003 &      0.004 &      0.060 &      0.109 &      0.168 &      0.076 &      0.137 &      0.224 &      0.096 &      0.190 & ... \\ 
   NGC3818 &  5 &   0 &      0.000 &      0.250 &      0.095 &      0.125 &      1.540 &      5.940 &      5.240 &      2.260 &      3.370 &      5.820 &      1.720 &      5.550 & ... \\ 
   NGC3818 &  5 &   1 &      0.115 &     22.200 &      0.003 &      0.004 &      0.061 &      0.110 &      0.171 &      0.077 &      0.139 &      0.228 &      0.097 &      0.194 & ... \\ 
   NGC3818 &  6 &   0 &      0.000 &      0.500 &      0.077 &      0.103 &      1.220 &      5.610 &      4.530 &      2.100 &      3.110 &      5.460 &      1.750 &      5.440 & ... \\ 
   NGC3818 &  6 &   1 &      0.190 &     22.200 &      0.007 &      0.008 &      0.133 &      0.230 &      0.412 &      0.190 &      0.335 &      0.549 &      0.190 &      0.470 & ... \\ 
    ...  & ... & ... & ... & ... & ... & ... & ... &... & ... & ... & ... & ... & ... & ... & ... \\
\hline
\label{table7}
\end{tabular}}

\medskip
First row: galaxy identification (col. 1), 
number of the aperture (col.~2) , flag indicating that index values 
are given (col.~3), radii delimited by the aperture 
(col.~4 and 5), indices for the apertures (col.~6 through 30). 
Second row: galaxy identification (col. 1), number of the 
aperture (col.~2) , flag indicating that error values 
are given (col.~3), luminosity weighted radius of the aperture 
(col.~4), adopted effective radius (col.~5),
errors on the indices (col.~6 through 30). 
The complete Table~8 is given in electronic form.
\end{table*}


\begin{table*}
\tiny{
\begin{tabular}{lrrrrrrrrrrrrrrrr}
 & & & & & & & & & & & & & & &  \\
\multicolumn{16}{c}{\bf Table 9} Fully corrected line--strength
indices for the gradients \\
\hline\hline
\multicolumn{1}{c}{galaxy} &
\multicolumn{1}{c}{iz} &
\multicolumn{1}{c}{ie} &
\multicolumn{1}{c}{r$_1$} &
\multicolumn{1}{c}{r$_2$} &
\multicolumn{1}{c}{CN$_1$} &
\multicolumn{1}{c}{CN$_2$} &
\multicolumn{1}{c}{Ca4227} &
\multicolumn{1}{c}{G4300} &
\multicolumn{1}{c}{Fe4383} &
\multicolumn{1}{c}{Ca4455} &
\multicolumn{1}{c}{Fe4531} &
\multicolumn{1}{c}{Fe4668} &
\multicolumn{1}{c}{H$\beta$} &
\multicolumn{1}{c}{Fe5015}&
\multicolumn{1}{c}{other}  \\
\multicolumn{1}{c}{galaxy} &
\multicolumn{1}{c}{iz} &
\multicolumn{1}{c}{ie} &
\multicolumn{1}{c}{r$_l$} &
\multicolumn{1}{c}{r$_e$} &
\multicolumn{1}{c}{eCN$_1$} &
\multicolumn{1}{c}{eCN$_2$} &
\multicolumn{1}{c}{eCa4227} &
\multicolumn{1}{c}{eG4300} &
\multicolumn{1}{c}{eFe4383} &
\multicolumn{1}{c}{eCa4455} &
\multicolumn{1}{c}{eFe4531} &
\multicolumn{1}{c}{eFe4668} &
\multicolumn{1}{c}{eH$\beta$}&
\multicolumn{1}{c}{eFe5015}&
\multicolumn{1}{c}{indices} \\
\hline
   NGC3607 &   0 &   0 &      0.000 &      0.063 &      0.144 &      0.182 &      1.640 &      5.950 &      6.060 &      2.410 &      3.700 &      8.390 &      1.510 &    5.750 & ...\\
   NGC3607 &   0 &   1 &      0.029 &     43.400 &      0.004 &      0.005 &      0.077 &      0.132 &      0.144 &      0.080 &      0.149 &      0.204 &      0.086 &    0.227 & ...\\
   NGC3607 &   1 &   0 &      0.063 &      0.125 &      0.106 &      0.135 &      1.570 &      5.800 &      5.780 &      2.310 &      3.940 &      7.450 &      1.480 &    5.760 & ...\\
   NGC3607 &   1 &   1 &      0.090 &     43.400 &      0.003 &      0.004 &      0.061 &      0.100 &      0.149 &      0.087 &      0.151 &      0.239 &      0.089 &    0.249 & ...\\ 
   NGC3607 &   2 &   0 &      0.125 &      0.250 &      0.079 &      0.110 &      1.410 &      5.730 &      5.570 &      2.340 &      3.580 &      6.170 &      1.580 &    5.580 & ...\\
   NGC3607 &   2 &   1 &      0.177 &     43.400 &      0.004 &      0.004 &      0.072 &      0.118 &      0.177 &      0.101 &      0.178 &      0.287 &      0.106 &    0.294 & ...\\
   NGC3607 &   3 &   0 &      0.250 &      0.500 &      0.082 &      0.111 &      1.690 &      5.670 &      5.790 &      2.330 &      3.350 &      6.020 &      1.550 &    5.550 & ...\\
   NGC3607 &   3 &   1 &      0.353 &     43.400 &      0.008 &      0.010 &      0.142 &      0.265 &      0.400 &      0.215 &      0.356 &      0.564 &      0.225 &    0.461 & ...\\
   NGC3818 &   0 &   0 &      0.000 &      0.063 &      0.158 &      0.194 &      1.670 &      6.120 &      5.860 &      2.350 &      3.580 &      7.630 &      1.610 &    5.870 & ...\\
   NGC3818 &   0 &   1 &      0.029 &     22.200 &      0.004 &      0.005 &      0.076 &      0.137 &      0.198 &      0.107 &      0.185 &      0.285 &      0.138 &    0.299 & ...\\
   NGC3818 &   1 &   0 &      0.063 &      0.125 &      0.122 &      0.153 &      1.700 &      6.090 &      5.760 &      2.310 &      3.900 &      6.970 &      1.630 &    5.760 & ...\\
   NGC3818 &   1 &   1 &      0.090 &     22.200 &      0.003 &      0.004 &      0.066 &      0.111 &      0.171 &      0.087 &      0.145 &      0.245 &      0.096 &    0.250 & ...\\
   NGC3818 &   2 &   0 &      0.125 &      0.250 &      0.077 &      0.103 &      1.480 &      5.780 &      4.910 &      2.220 &      2.940 &      5.140 &      1.950 &    5.450 & ...\\
   NGC3818 &   2 &   1 &      0.174 &     22.200 &      0.003 &      0.004 &      0.069 &      0.116 &      0.179 &      0.090 &      0.148 &      0.242 &      0.094 &    0.237 & ...\\
   NGC3818 &   3 &   0 &      0.250 &      0.500 &      0.037 &      0.065 &      0.904 &      5.380 &      3.500 &      1.990 &      3.080 &      4.040 &      2.060 &    5.480 & ...\\
   NGC3818 &   3 &   1 &      0.348 &     22.200 &      0.006 &      0.007 &      0.123 &      0.199 &      0.326 &      0.165 &      0.268 &      0.458 &      0.168 &    0.422 & ...\\
    ... & ... & ... & ... & ... & ... & ... &... & ... & ... & ... & ... & ... & ... & ...  & ...  \\									   
\hline
\label{table8}
\end{tabular}}
\medskip

First row: galaxy identification (col. 1), 
number of the gradient (col.~2) , flag indicating that index values 
are given (col.~3), radii delimited by the gradient 
(col.~4 and 5), indices for the gradients (col.~6 through 30). 
Second row: galaxy identification (col. 1), number of the 
gradient (col.~2) , flag indicating that error values 
are given (col.~3), luminosity weighted radius of the gradient 
(col.~4), adopted effective radius (col.~5),
errors on the indices (col.~6 through 30). 
The complete Table~9 is given in electronic form.

\end{table*}


\section{Measurements of line-strength indices and transformation into the Lick IDS System}

The procedure adopted to extract line--strength indices 
in the Lick IDS System from the galaxy spectra
has been widely described in Paper~I.
Here we summarize the basic steps. 

Since our spectral resolution (FWHM $\sim$ 7.6 \AA\  at $\sim$ 5000 \AA) 
on the entire spectrum  is slightly better 
than the wavelength-dependent 
resolution of the Lick IDS
system (see Worthey \& Ottaviani \cite{OW97}), we have
degraded our data convolving each spectrum
(apertures and gradients) with a wavelength-dependent gaussian kernel 
(see equation (5) of Paper~I).
On the smoothed spectra we have measured 25 line-strength indices: 
21 of the original Lick IDS system 
(see Table~2 in Trager et al. \cite{Tra98} for the index 
bandpass definitions) plus 4 higher order Balmer lines later introduced 
by Worthey \& Ottaviani  (\cite{OW97}) 
(see their Table~1 for the index definitions).

\subsection{Correction for velocity dispersion}

The observed spectrum of a galaxy can be regarded as a stellar spectrum
convolved with the radial velocity distribution of its stellar population.  
Therefore spectral
features in a galactic spectrum are not the simple sum of its corresponding
stellar spectra, because of the stellar motions.  To measure the stellar
composition of galaxies, we need to correct index measurements for the
effects of the galaxy velocity dispersion (see e.g. Gonz\' alez \cite{G93}
(hereafter G93), Trager et al. (\cite{Tra98}); Longhetti et al. (\cite{L98a})). 

To this purpose in Paper~I we have 
selected stars with spectral type between G8III and K2III among 
the Lick stars observed together with the galaxies, and have convolved
their spectra with gaussian curves of various widths in order 
to simulate different galactic velocity dispersions.
On each convolved spectrum we have then measured 25 Lick-indices and
have derived the fractional index variations as function of
the velocity dispersion $\sigma$:

\begin{equation}
\label{eq1}
R_{i,j,\sigma}= (\frac{EW_{i,j,\sigma}-EW_{i,j,0}}{EW_{i,j,0}}) 
\end{equation}

where $R_{i,j,\sigma}$ is the fractional variation of the i-th index
measured for the j-th star at the velocity dispersion $\sigma$, while
$EW_{i,j,\sigma}$ and $EW_{i,j,0}$ are the index values at the velocity 
dispersion $\sigma$ and at zero velocity dispersion respectively.
To correct the galactic indices we used a mean correction
derived through an average on the $R_{i,j,\sigma}$ values
(see Section 4.2 of Paper~I for details).

\begin{figure}
\psfig{figure=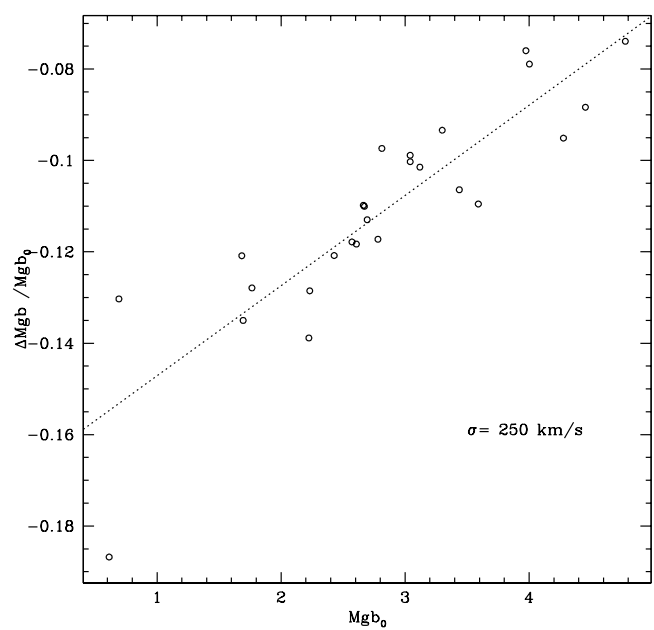,width=9cm,clip=}
\caption{Fractional variation of the Mgb index for a velocity
dispersion $\sigma=$250  km~s$^{-1}$  as function of  Mgb$_0$,
the index at zero velocity dispersion. 
The dotted line is the least squares
fit to the data. } 
\label{fig3}
\end{figure}

The use of an average correction is a reasonable choice 
as long as the fractional index change $R_{i,j,\sigma}$ does
not depend on the index value at fixed velocity dispersion.
On the contrary we have verified that in some cases the fractional 
index change presents a correlation with
the line strength index, as shown in Figure \ref{fig3} for  Mgb, 
and the use of an average velocity dispersion correction may be not the best 
strategy to adopt.
For this reason we have implemented a new procedure to correct 
the total sample (50+15 galaxies) for velocity dispersion.

We consider the individual fractional index variations
 $R_{i,j,\sigma}$  and associated line--strength indices $EW_{i,j,\sigma}$
derived for each star at different velocity dispersions. 
Then for each galactic aperture or gradient
the final index correction $R_{i,\sigma}$ is derived by interpolating 
the  $R_{i,j,\sigma}$  values at 
the galactic velocity dispersion $\sigma$ and at the
uncorrected index value $EW_{i,old}$.

For each gradient and aperture the velocity dispersion 
corrections are computed on the basis
of the $\sigma$ values listed in Table~4.
The tabulated values characterize the trend of each galaxy 
velocity dispersion curve. 
For galaxies having only the central (r$_e$/8) 
estimate of $\sigma$ we adopt this value also
for the correction of the indices at larger radii (the tables of indices
uncorrected for velocity dispersion are available,  under request,  
to authors). 

The new index corrected for the effect of velocity dispersion is computed 
in the following way:

\begin{equation}
\label{eq2}
EW_{i,new}=EW_{i,old}/(1+R_{i,\sigma})
\end{equation} 

\begin{figure*}
\resizebox{8.5cm}{!}
{\psfig{figure=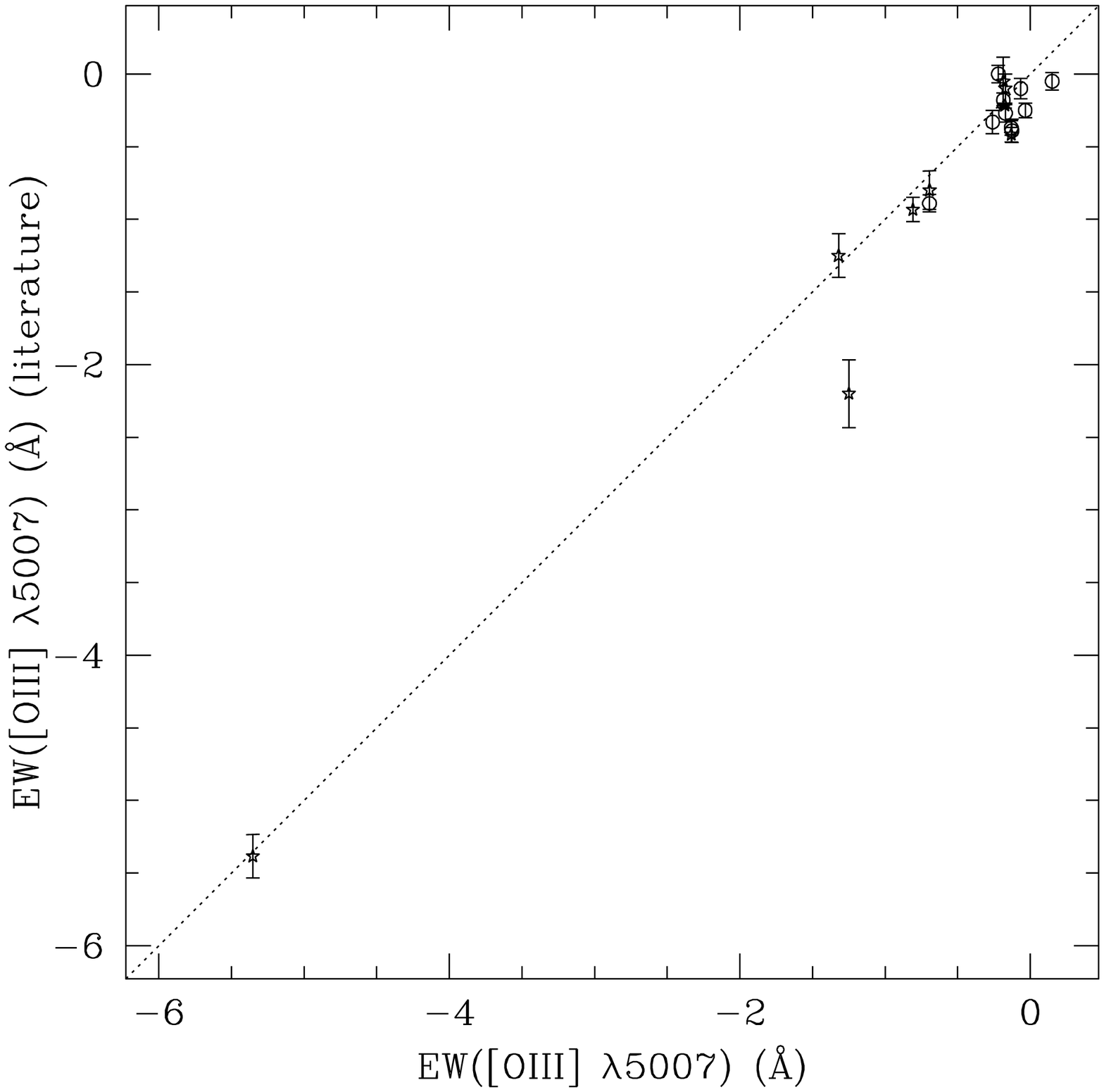,width=16cm,clip=}}
\resizebox{8.5cm}{!}
{\psfig{figure=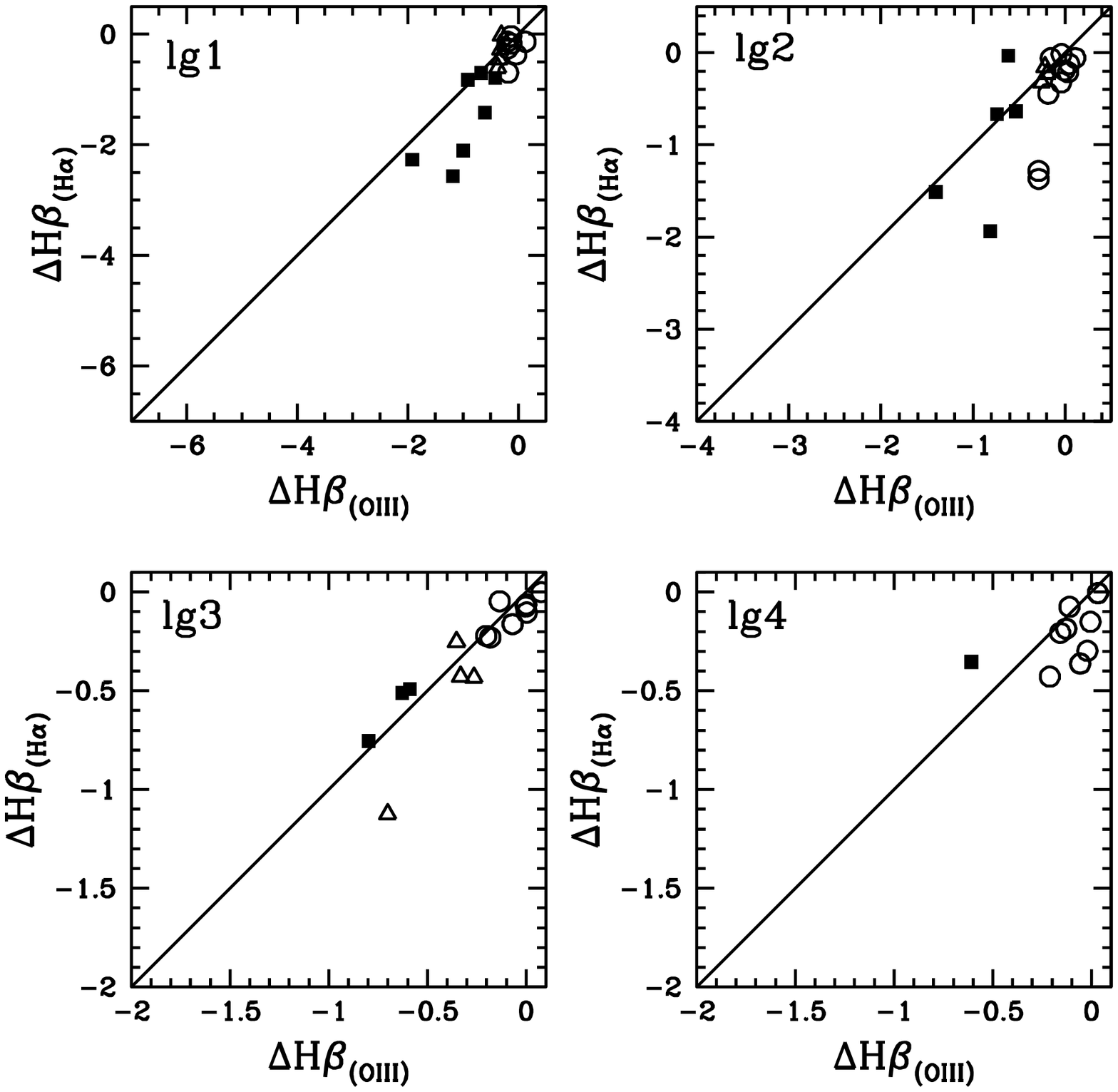,width=16cm,clip=} }
\caption{{\bf Left panel:}
comparison between our ([OIII]$\lambda 5007$) emission estimates with 
the measures of Trager et al. \cite{Tra00} (open circles) 
and Denicol{\' o} et al. \cite{Deni05} (open stars)
within an $r_e/8$ aperture for the galaxies in common.
The dotted line marks the one-to-one relation.
{\bf Right panel:}
comparison between the  H$\beta$ 
emission estimates  derived from the  ([OIII]$\lambda 5007$)  
and from the {H$\alpha$} 
emission lines respectively (see Section 4.2).  
The solid line is the one-to-one relation. 
The comparison is shown in the four regions sampled by
the linear gradients (legend: lg1 (0 $\leq$ r $\leq$r$_e$/16("nuclear")), lg2
(r$_{e}$/16 $\leq$ r $\leq$r$_e$/8),  
lg3 (r$_{e}$/8 $\leq$ r $\leq$r$_e$/4) and
lg4 (r$_{e}$/4 $\leq$ r $\leq$r$_e$/2). Open circles indicate galaxies which
O[III] emission is detected under 1$\sigma$ level, triangles and full squares
between 1 and 2$\sigma$ levels and above 2$\sigma$ level 
respectively (see text).}
\label{fig4}
\end{figure*}

\subsection{Correction for emission}

The presence of emission lines affects the measure of some line--strength
indices. In particular, the H$\beta$ absorption strength of 
the underlying stellar
population could be contaminated by a significant infilling due to presence of
the H$\beta$ emission component. The adopted correction procedure  
is detailed in Paper~I.

We recall here that the correction of 
the galaxy line--strength indices is performed through the measure of
the H$\beta$ emission component from the [OIII] emission, according to 
the relation proposed by G93:
EW(H$\beta_{em}$)/EW([OIII]$\lambda 5007) = 0.7$.
 
The ([OIII]$\lambda 5007$) emission is derived adopting
a template spectrum for the underlying
stellar population and measuring in the galaxy spectrum
the flux in excess with respect to the template in a band centered 
around 5007 \AA.
Following Goudfrooij's (\cite{Gou98}) suggestion to use the spectrum
of an elliptical galaxy to this purpose, we looked among
our observed galaxies for those lacking evidence of emission lines or dust 
in their spectra and images, and selected  NGC~1426 as an old
population template.
Our choice of a galaxy rather than a stellar template,
as done instead by G93,
is fully motivated in Paper~I, to which we refer for details.

Once the template spectrum has been degraded to the velocity dispersion
of the galaxy region under exam and normalized to the galaxy continuum,
the galaxy and template fluxes are measured within 
a bandpass (4996.85 - 5016.85) centered at 5007\AA, while the pseduo-continuum 
is defined by a blue
(4885.00 - 4935.00) and a red (5030.00 - 5070.00) bandpass (G93).
The emitted ([OIII]$\lambda 5007$) flux is then derived according to the
equation:

\begin{equation}
\label{eq3} 
EW_{em}=\int_{\lambda_1}^{\lambda_2} \frac{F_R-F_{temp}}{F_C} d \lambda
\end{equation}

where $F_R$, $F_{temp}$ and $F_C$ are  the galaxy, the template and 
the continuum fluxes respectively, while $\lambda_1$ and $\lambda_2$ 
define the central band.
According to this definition, detected emissions result as negative EWs. 
Considering the ([OIII]$\lambda 5007$) emissions detected above 1 $\sigma$
(the variance of the spectrum), we derived the EW of the H$\beta$ emission 
according to the G93 relation. 

We have compared our derived [OIII] emissions with literature estimates.
There are 10 galaxies in common between our complete sample of (50 + 15) 
galaxies and the 
sample of G93, for which Trager et al. \cite{Tra00} 
give [OIII] emission  (NGC~1453, NGC~3818, NGC~4374, NGC~4552, 
NGC~4697, NGC~5638, NGC~5812, NGC~5813, NGC~5831, NGC~5846); 
among the 15 galaxies in
common with Denicol{\' o} et al. \cite{Deni05}, 
[OIII] emission is measured for 9 objects 
(NGC~1052, NGC~1453, NGC~2911, NGC~2974, NGC~3607, NGC~4374, 
NGC~5363, NGC~5813,  NGC~5831).
The comparison is shown in the left panel of Figure \ref{fig4} for an 
$r_e/8$ aperture: open circles 
and open stars denote respectively the measures of 
Trager et al. \cite{Tra00} and Denicol{\' o}  et al. \cite{Deni05}, while
the dotted line marks the one-to-one relation.
For the majority of the galaxies our emission 
estimates are consistent with literature measures within the errors.
The largest deviation from the one-to-one relation is observed for
NGC~2911, for which we measure an [OIII] emission
of -1.25,  whereas in the literature we find  -2.2 $\pm$ 0.23.

As done for the original 50 galaxies of the sample of Paper~I,
we have measured  H$\alpha$ emissions for the 18 galaxies of this paper 
as well, and have derived the H$\beta$ emission
according to the relation  $F_{H\beta} =  1/2.86 F_{H\alpha}$ (see e.g. 
Osterbrock~\cite{Osterb89}).
  
The de-blending of the H$\alpha$ emission at $\lambda 6563$ from the
([NII]$\lambda$ 6548,6584) emission lines is performed by fitting
each galaxy spectrum (apertures and gradients) with a model  
resulting from the sum of our old population template galaxy and
three gaussian curves of arbitrary  widths and amplitudes 
(see Figure~4 of Paper~I as an example). 
Once derived the emitted  flux $F_{H\beta}$ 
from $F_{H\alpha}$, we computed the pseudo-continuum in 
H$\beta$ according to the bandpass definition of Trager et al. 
(\cite{Tra98}) and used it to transform 
the emitted flux $F_{H\beta}$ into EW.
In the right panel of Figure~\ref{fig4} we plot the comparison  
between the H$\beta$ emission estimates derived 
from the ([OIII]$\lambda 5007$ ) and the  $H\alpha$ lines
respectively.

Finally, the corrected  H$\beta_{corr}$  index  
is computed from the raw value 
H$\beta_{raw}$ according to the formula 
EW (H$\beta_{corr}$) = EW(H$\beta_{raw}$) - H $\beta_{em}$, 
where  H $\beta_{em}$  is obtained from  the [OIII] emission
as previously described.
The right panel of Figure~\ref{fig4} shows that 
this latter estimate is {\it statistically similar} to that 
obtained from the  H$\alpha$ emission, although
the use of the H$\alpha$ line for emission correction will be widely discussed 
in a forthcoming paper.

Table~5 and Table~6 report the values of the   H$\beta$ correction for the
apertures and gradients derived from O[III] and H$\alpha$ (complete tables are 
given in electronic form).

\subsection{Lick IDS Standard Stars}

After the indices have been homogenized to the Lick IDS 
wavelength dependent 
resolution, corrected for emission and velocity dispersion, they need a final
correction to be transformed into the Lick system.
The procedure that we adopted is described in  
Worthey \& Ottaviani (\cite{OW97}) 
and consists in observing, contemporary to the galaxies, a sample 
of standard stars of different spectral types common to the Lick library.
In Paper~I we have derived
line strength indices for 17 observed standard stars 
and have compared our measures with the values reported 
by Worthey et al. \cite{Wor94} in order to derive the 
transformations into the Lick system.
We report in Table~7 
the parameters $\alpha$ and $\beta$ of the linear transformation 
$EW_{Lick} = \beta + \alpha \times EW_{our}$ derived in Paper~I, 
 where $EW_{our}$  and $EW_{Lick}$ are the raw and the Lick 
indices respectively. 
The same parameters are used in this paper to perform the transformation 
into the Lick system.
We notice that for the majority 
of the indices $\alpha$ is very close to 1 and only a 
zero-point correction is required
(see also Puzia et al. \cite{Puzia2002}), although serious 
deviations from the one-to-one
relation are observed for Ca4227, Ca4455 and Fe4531.

\subsection{Estimate of index errors}

In order to obtain the errors on each measured index we have used the following
procedure. Starting from a given extracted spectrum 
(aperture or gradient at different
galactocentric distances), we have generated a  set of
1000 Monte Carlo random modifications, by adding a wavelength dependent 
Poissonian fluctuation from the corresponding spectral noise, 
$\sigma(\lambda)$.
Then, for each spectrum, we have estimated the moments of the distributions of
the 1000 different realizations of its indices.


\begin{table}
\small{
\begin{tabular}{lcrrr}
& &  & &\\
\multicolumn{5}{c}{\bf Table 9. ~~~ Comparison with the literature} \\
\hline\hline
\multicolumn{1}{c}{index}
&\multicolumn{1}{c}{N$_{gal}$}
& \multicolumn{1}{c}{offset}
& \multicolumn{1}{c}{dispersion}
& \multicolumn{1}{c}{units}
 \\
 \hline
 & & G93 + Long98 & & \\
\hline
        H$\beta$       & 14 &   0.028 &   0.614 & \AA  \\
       Mg2       & 14 &  -0.012 &   0.021 & mag  \\
       Mgb       & 14 &  -0.136 &   0.375 &\AA   \\
    Fe5270       & 14 &   0.069 &   0.295 &\AA   \\
    Fe5335       & 14 &  -0.169 &   0.446 &\AA   \\
       Mg1       & 14 &  -0.013 &   0.021 & mag  \\
 \hline                                     
 & & Trager et al. (1998) & & \\            
\hline
        H$\beta$       & 28 &   0.004 &   0.375 &\AA  \\
       Mg2       & 28 &  -0.016 &   0.026 &mag  \\
       Mgb       & 28 &  -0.203 &   0.447 &\AA   \\
    Fe5270       & 29 &  -0.094 &   0.506 &\AA   \\
    Fe5335       & 29 &  -0.241 &   0.504 &\AA   \\
       Mg1       & 28 &  -0.015 &   0.021 &mag  \\
     G4300       & 29 &  0.153  &   0.887 &\AA    \\
     Ca4227      & 29 &  0.527  &   0.797 &\AA    \\

 \hline
 & & Beuing et al. (2002)& & \\
\hline
        H$\beta$       & 12 &  -0.043 &   0.273 &\AA  \\
       Mg2       & 12 &   0.015 &   0.024 &mag  \\
       Mgb       & 12 &   0.188 &   0.610 &\AA   \\
    Fe5270       & 12 &   0.277 &   0.603 &\AA   \\
    Fe5335       & 12 &   0.076 &   0.405 &\AA   \\
       Mg1       & 12 &   0.008 &   0.018 &mag  \\
 \hline                                     
 & & Denicol{\' o} et al.(2005)& & \\                
\hline
      H$\beta$   & 15 &   0.574 &   1.291 &\AA\\ 
       Mg2       & 15 &   0.001 &   0.021 &mag  \\
       Mgb       & 15 &  -0.060 &   0.513 &\AA   \\
    Fe5270       & 15 &   0.117 &   0.455 &\AA   \\
    Fe5335       & 15 &  -0.052 &   0.333 &\AA   \\
       Mg1       & 15 &  -0.003 &   0.018 &mag  \\
    \hline\hline                           
\end{tabular}}                             
\label{table9}

\medskip
Offsets and dispersions of the residuals between our data and the literature. 
Dispersions are 1 $\sigma$ scatter of the residuals.
Beuing et al. (2002) do not compute G4300 and Ca4227 indices. For G4300 and Ca4227 indices
Gonzalez (1993) and Longhetti et al. (1998) used a (slightly) different definition than 
Trager et al. (1998)  and consequently comparisons are not reported in the table. The 
comparison for H$\beta$ index between our data and those of Trager et al. (1998) and
Beuing et al. (2002) are made using uncorrected data. 
\end{table}



\section{Results} 

For each galaxy of the sample, 25 Lick indices obtained for the 7 apertures 
and the 4 gradients are provided in electronic form  
with the format shown in Tables~8 and 9 respectively. 
We provide also the indices of the original set of
Paper~I corrected for velocity dispersion according
to the new procedure implemented in this paper.
The structure of the above tables is the following: 
each aperture (or gradient) is described
by two rows. In the first row: col.~1 gives the galaxy identification, 
col.~2 the number of the aperture, col.~3 is a flag 
(0 stands for values of indices),
col.~4 and 5 give the radii delimited by the aperture, 
from col.~6 to 30 individual indices are given. 
In the second row: col.~1 gives the galaxy identification, col.~2 the number
of the aperture, col.~3 is a flag (1 stands for error of the indices),
col.~4 and 5 give the luminosity weighted radius of the aperture and the
adopted equivalent radius, from col.~6 to 30 the errors of the indices 
are given.
In electronic form are also available, under request to the authors, 
the tables of the raw indices  (before corrections for 
velocity dispersion, emission and transformation into the Lick system)
as well as the fully calibrated spectra 
(apertures and gradients) in digital form for each galaxy.

\begin{figure}
\psfig{figure=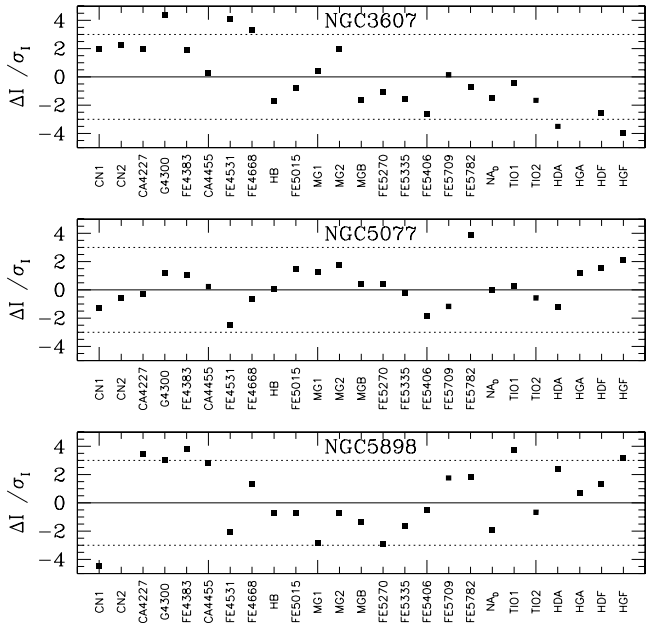,width=9cm,clip=}
\caption{Comparison of the indices presented Paper~I
with the measures of this paper for the  
three galaxies in common 
(NGC~3607, NGC~5077 and NGC~5898) within an  $r_e/8$ aperture.
The index differences are plotted in units of the standard error.
The dotted lines delimit the region where $\Delta I/I_{\sigma}<3$}. 
\label{fig5}
\end{figure}

In Figure \ref{fig5} we have compared the indices from Paper~I 
and from this paper for the three galaxies in common 
(NGC~3607, NGC~5077 and NGC~5898) within an  $r_e/8$ aperture.
The figure shows that the two index measures are generally in agreement
within 3 $\sigma$.

\begin{figure*}
\psfig{figure=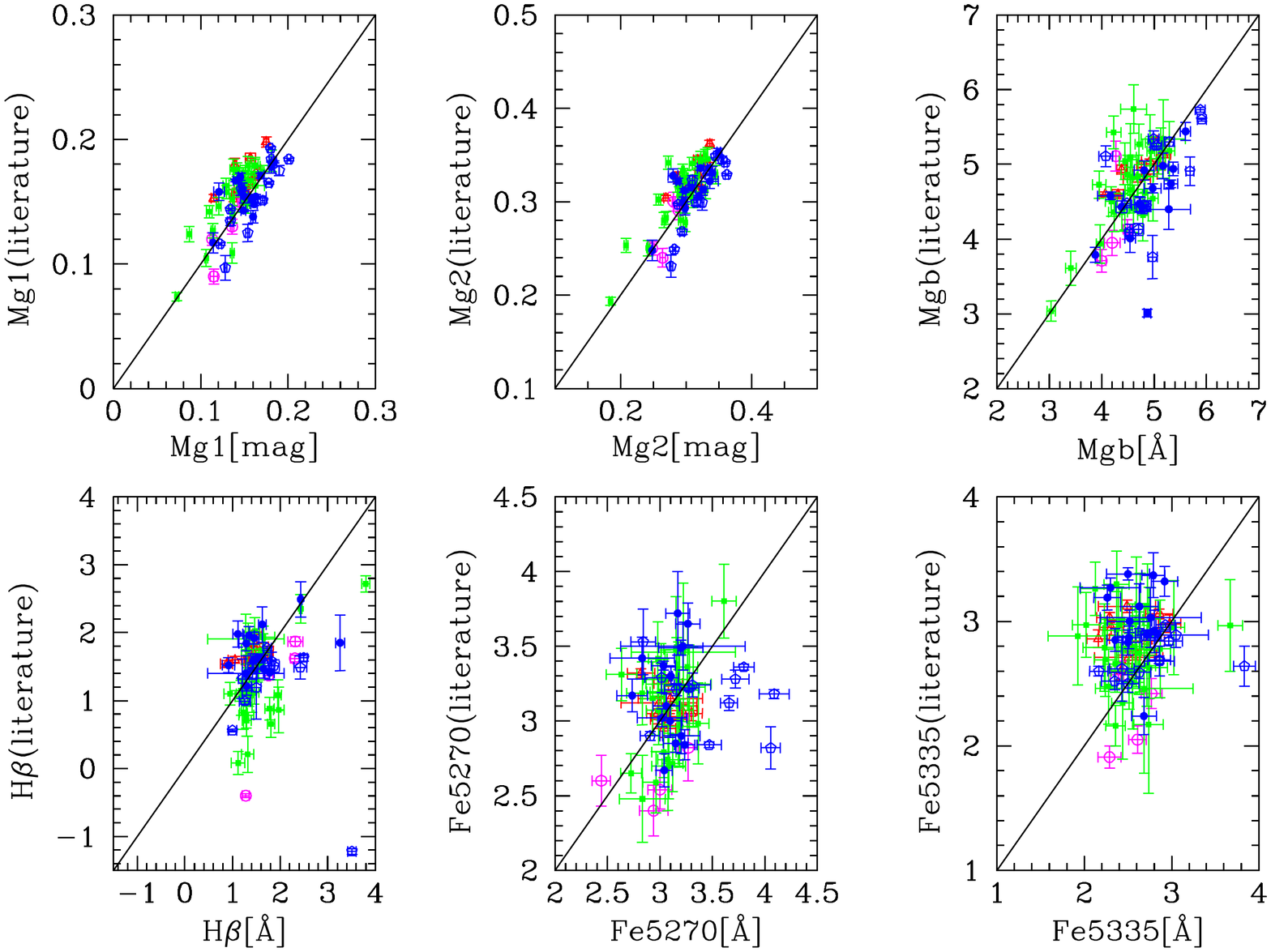,width=16cm,angle=-90,clip=}
\caption {Comparison of index measurements of Gonz\' alez (1993: open
triangles), Trager (1998: full squares), 
Longhetti et al. (1998: open circles),  Beuing et al. 
(2002: open pentagons) and Denicol{\' o} et al. 
(2005: full circles) 
 with our data. Solid lines mark the one-to-one
relation. Table~9 summarizes the results of the comparison. } 
\label{fig6}
\end{figure*}

The set of on-line indices in the literature available
for a comparison is quite heterogeneous since indices
are measured within different apertures.
There are ten galaxies in common between the {\it total sample } and 
the G93 sample, namely NGC~1453, NGC~4552, NGC~5846, 
NGC~3818, NGC~4374, NGC~4697, NGC~5638, NGC~5812, NGC~5813 and NGC~5831.
Four galaxies belong to the Longhetti et al. (\cite{L98a}) sample
(NGC~1553, NGC~6958, NGC~7135 and NGC~6776).
 Twenty nine galaxies are in the sample of Trager et al. (\cite{Tra98})
 (NGC~128, NGC~777, NGC~1052, NGC~1209, NGC~1380, NGC~1407, 
 NGC~1426, NGC~1453, NGC~1521, NGC~2749, NGC~2962, NGC~2974,
 NGC~3489, NGC~3607, NGC~3962, NGC~4552, NGC~4636, NGC~5077, 
 NGC~5328, NGC~7332, NGC~7377, NGC~3818, NGC~4374, NGC~4697,
 NGC~5044, NGC~5638, NGC~5812, NGC~5813, NGC~5831).
Twelve galaxies are in the sample published by Beuing et al.  \cite{Beu02},
(IC~1459, IC~2006, NGC~1052, NGC~1209, NGC~1407, NGC~1553, NGC~5898, 
NGC~6868, NGC~6958, NGC~7007, NGC~7192 and NGC~5812).
Finally  fifteen galaxies are in common with the sample recently published
by Denicol{\' o} et al. \cite{Deni05} (NGC~777, NGC~1052, NGC~1407, 
NGC~1453, NGC~2911, NGC~2974, NGC~3607, NGC~4374, NGC~5363, NGC~5638,  
NGC~5812, NGC~5813, NGC~5831, NGC~5846, NGC~7332).
The comparison with the literature for the {\it total sample }
is presented in Figure \ref{fig6}. 
In detail: (1) with Longhetti et al. (\cite{L98a}) the comparison is 
made with indices computed 
on the aperture of 2.5\arcsec radius; (2) with Gonzalez (\cite{G93}) on r$_e/8$
aperture and (3) with Trager et al. (\cite{Tra98})  with indices computed 
within standard apertures; (4) with Beuing et al. \cite{Beu02} with indices 
computed on the aperture with radius r$_{e}$/10, 
taking into account that these  authors did not correct H$\beta$  
for emission infilling; (5) with  Denicol{\' o} et al. \cite{Deni05} 
on r$_e/8$ aperture. 

Table~9 presents a summary of the comparison with the literature.
Both the offset and the dispersion for the various indices in the table are
comparable (or better) of those obtained on the same indices by 
Puzia et al.~\cite{Puzia2002} in their spectroscopic study 
of globular clusters.    
Notice that the discrepancy with Denicol{\' o} et al. \cite{Deni05} regarding 
the H$\beta$ index reduces to an offset of 0.090  and a dispersion 
of  0.404 when the comparison is done with the index values not 
corrected for emission.

\begin{figure}
\psfig{figure=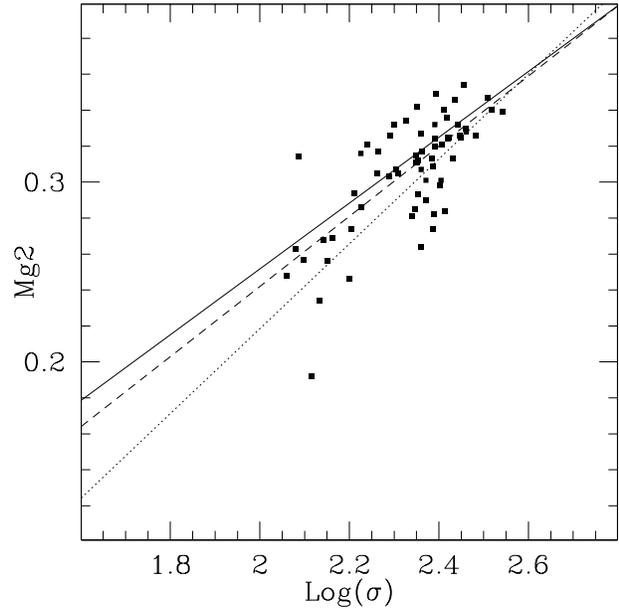,width=8.5cm,clip=}
\caption{ Mg$_2$ versus $\sigma$ relation. 
We plot our fully corrected Mg$_2$ line-strength indices 
within the SLOAN aperture (r=1.5\arcsec) versus the corresponding
velocity dispersion values. 
The solid line is the least-squares fit obtained by  
Trager et al. ~(\cite{Tra98}). The dotted line represents 
the least-squares fit obtained by Bernardi et al. (1998) for the field 
sample of 631 galaxies, while the long-dashed line is our fit
(value at $\sigma_{300~km~s^{-1}}$ = 0.335,
slope of relation =  0.195) to the present data.}
\label{fig7}
\end{figure}

The Mg$_2$ vs. $\sigma$ relation is plotted in Figure \ref{fig7}
for the {\it total sample} of (50+15) galaxies.
For a comparison with the SLOAN data, 
we plot our Mg$_2$ values at the r=1.5\arcsec
versus the corresponding velocity dispersion values. The
dotted line is the least-squares fit obtained by Bernardi et al.
(\cite{Ber98}) on their sample of 631 field early-type galaxies, while the
solid line represents the Trager et al. (\cite{Tra98})  fit.
We recall that the
Bernardi et al. (\cite{Ber98}) fit is performed on Mg2 index values computed adopting
the Lick bandpass definitions, but without transformation 
into the Lick IDS system. The
long-dashed line marks our least-squares fit for the total sample: 
we observe that our slope is well consistent with that of 
Bernardi et al. (\cite{Ber98}). 

In Figures 8-10 we show as examples the trend with radius of the
 Mg$_2$, Fe(5335) and H$\beta$ indices for the 18 galaxies of the sample 
(apertures are marked with open squares, gradients with dots).  


\section{Summary}

This {\it addendum} is dedicated as Paper~I to the characterization of 
the underlying stellar population in early-type galaxies with emission 
lines through the preparation of a data base of their
line-strength indices in the Lick IDS system.
The indices are measured on the galaxy spectra and then 
corrected for several effects, more specifically 
infilling by emission, velocity dispersion and 
transformation into the Lick IDS system. 
This paper enlarges the data-set
of line-strength indices of 50 early-type galaxies already 
presented in Paper~I by analyzing  18 additional early-type galaxies 
(three galaxies, namely NGC~3607, NGC~5077 and 
NGC~5898 have been already presented in the previous set).

For each object we have extracted 7 luminosity weighted apertures 
(with radii: 1.5\arcsec, 2.5\arcsec, 10\arcsec,  r$_e$/10, r$_e$/8, 
r$_e$/4 and r$_e$/2) corrected for the galaxy ellipticity and  
4 gradients (0 $\leq$ r $\leq$r$_e$/16, r$_{e}$/16 $\leq$ r $\leq$r$_e$/8,
r$_{e}$/8 $\leq$ r $\leq$r$_e$/4 and r$_{e}$/4 $\leq$ r $\leq$r$_e$/2). For
each aperture and gradient we have measured 25 line--strength 
indices: 21 of the original set
defined by the Lick IDS ``standard'' system (Trager et al. \cite{Tra98}) and 4 
later introduced by Worthey \& Ottaviani  (\cite{OW97}).

Line-strength indices, in particular those used to build the 
classic  H$\beta$--[MgFe] plane, have been compared with the literature. 
A direct comparison was made with Gonz\' alez (\cite{G93}), 
Longhetti et al. (\cite{L98a}), Beuing et al. (\cite{Beu02}), 
Trager et al. (\cite{Tra98})  and Denicol{\' o} et al. \cite{Deni05}, 
showing the reliability of our measures.


\begin{figure*}
\psfig{figure=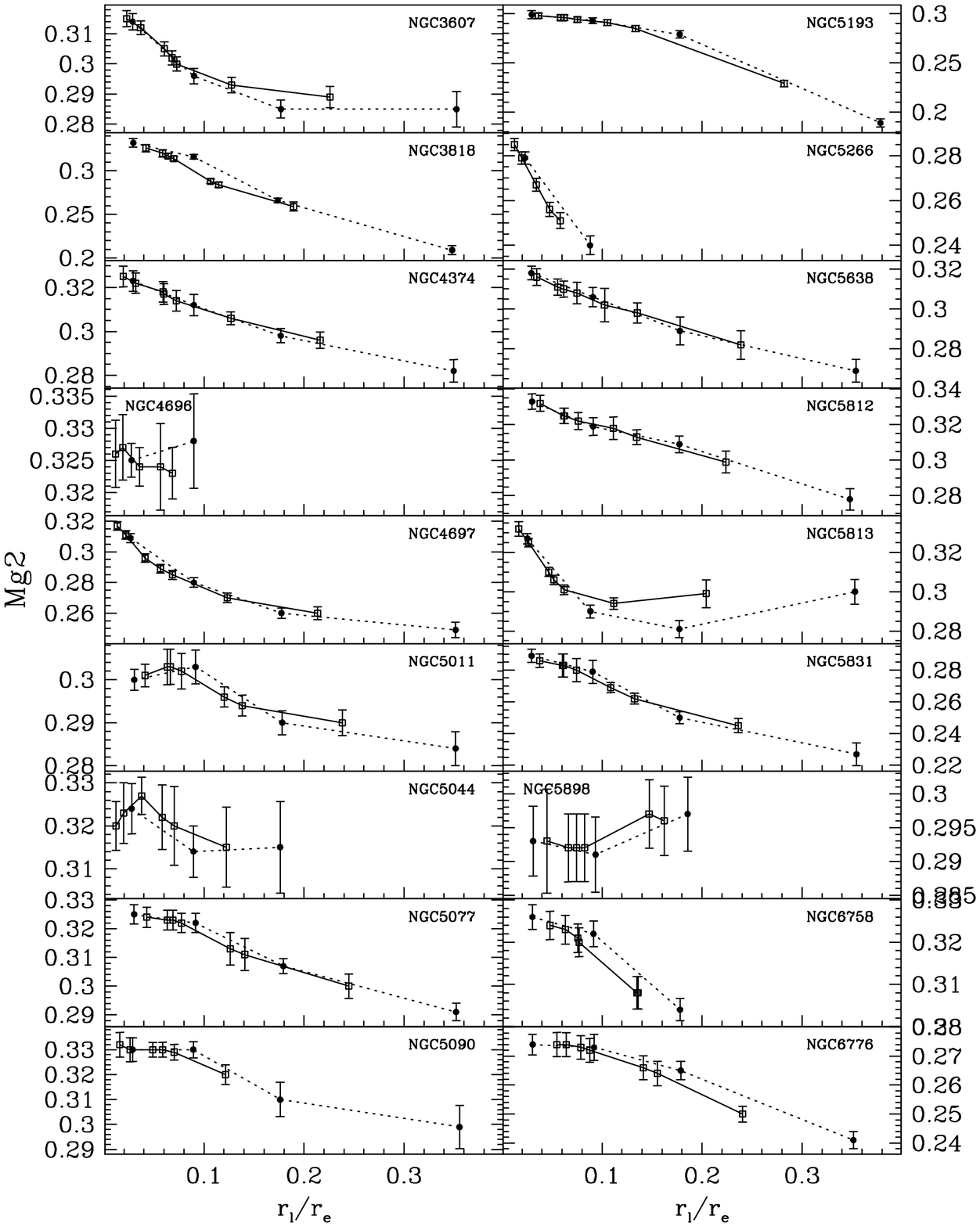,width=16cm,clip=}
\caption{Fully corrected Mg$_2$ line-strength index as function of 
the luminosity weighted radius normalized to the galaxy equivalent 
radius R$_e$. Apertures are indicated with open squares, 
while gradients are indicated with full dots.} 
\label{fig8}
\end{figure*}

\begin{figure*}
\psfig{figure=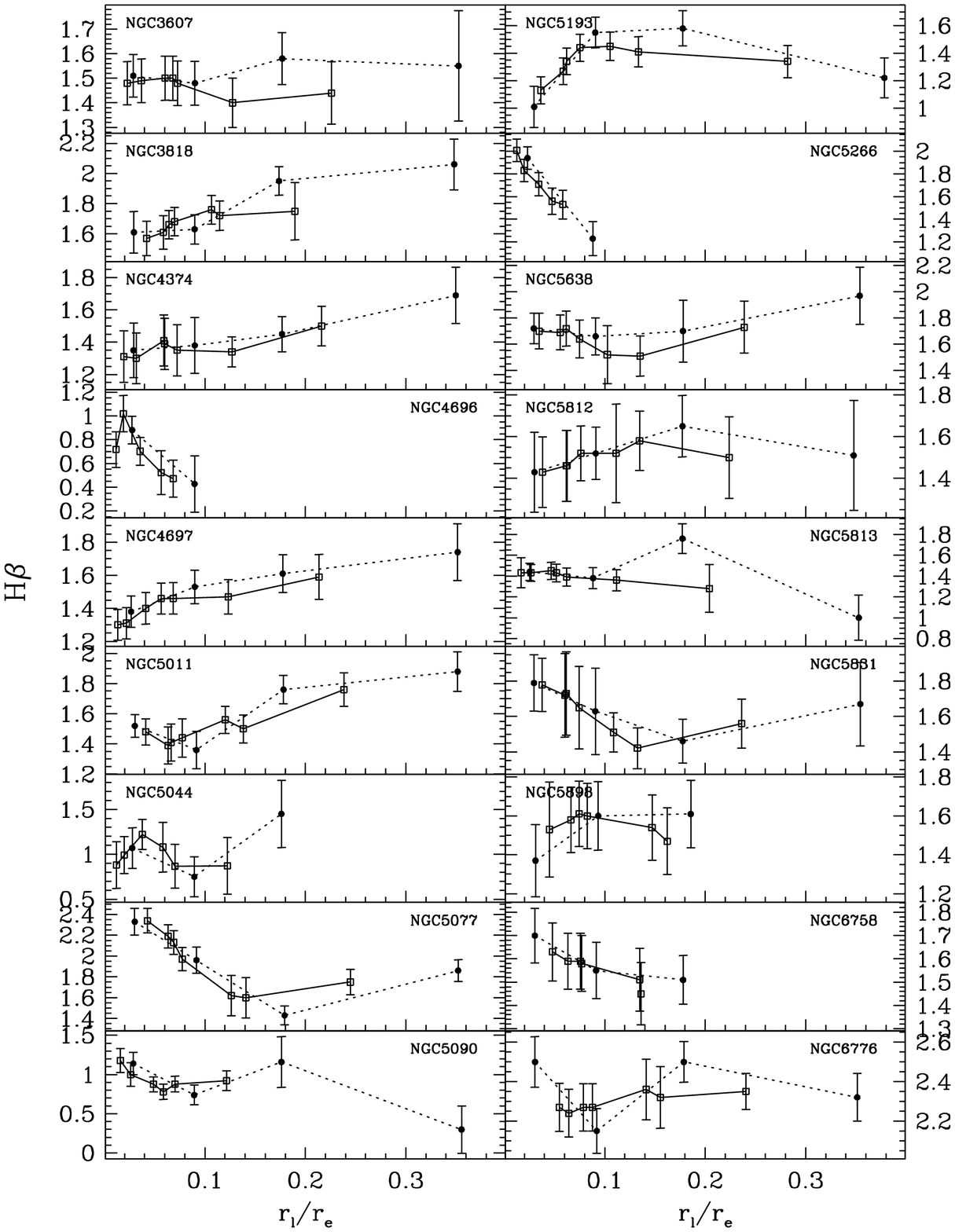,width=16cm,clip=}
\caption{Fully corrected H$\beta$ line-strength index as function 
of the luminosity weighted radius normalized to the galaxy equivalent 
radius R$_e$. Apertures are
indicated with open squares, while gradients are indicated with full dots.} 
\label{fig9}
\end{figure*}

\begin{figure*}
\psfig{figure=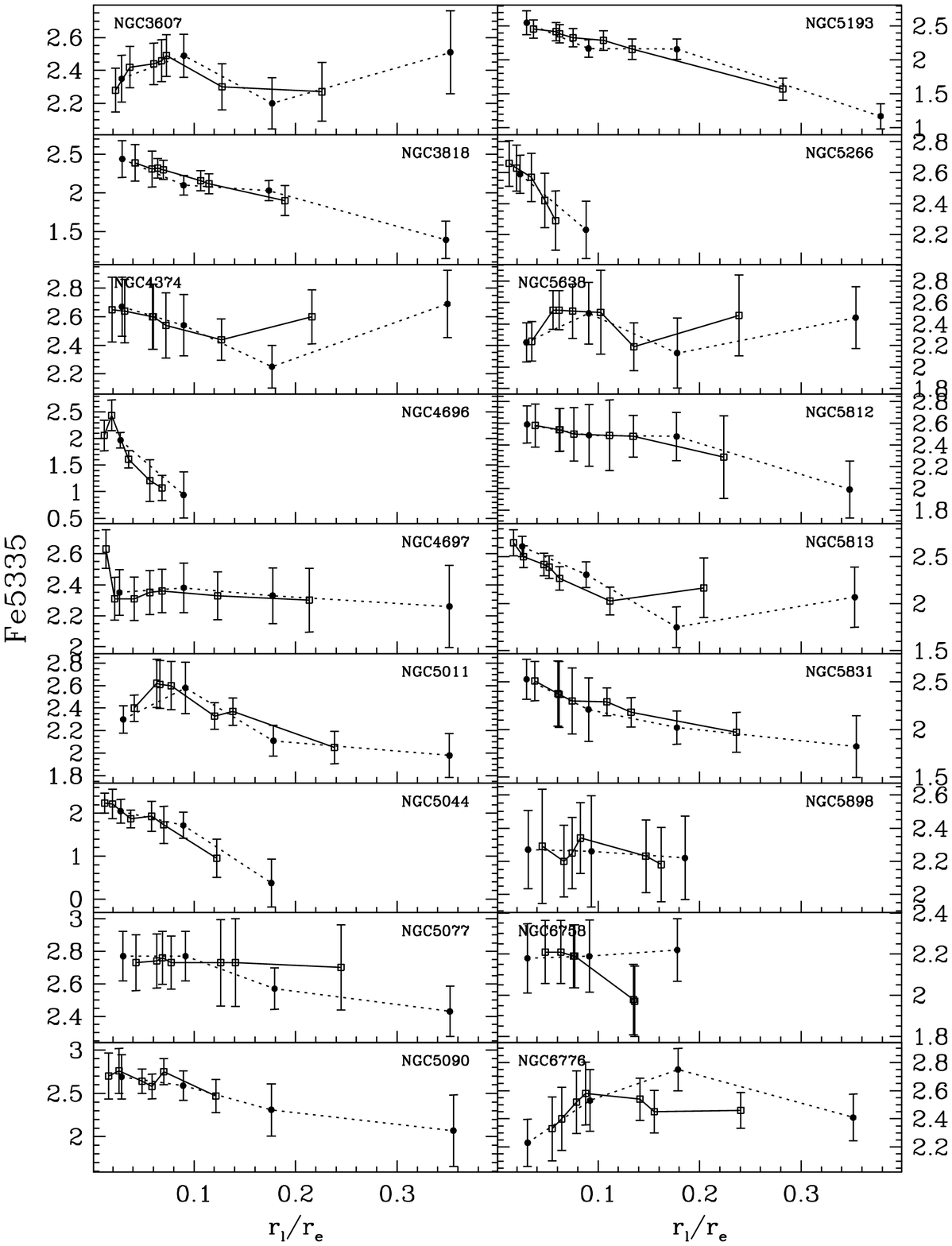,width=16cm,clip=}
\caption{Fully corrected Fe5335 line-strength index as function 
of the luminosity
weighted radius normalized to the galaxy equivalent radius R$_e$. Apertures are
indicated with open squares, while gradients are indicated with full dots.} 
\label{fig10}
\end{figure*}

\begin{acknowledgements} RR acknowledges  the partial support of the ASI
(contract I/R/037/01). WWZ acknowledges the support of the Austrian Science Fund 
(project P14783) and of the Bundesministerium f\"ur Bildung, Wissenschaft und Kultur.
This research has made use of the NASA/IPAC Extragalactic Database (NED) which is operated by the Jet Propulsion Laboratory, California Institute of
Technology, under contract with the National Aeronautics and Space
Administration.
\end{acknowledgements}

\appendix
\section{Relevant notes on individual galaxies from the literature}

We report below some studies relevant to the present investigation
performed in the recent literature. Our attention is addressed to the properties 
of the ionized gas with respect to the bulk of the stellar component and to the
cold or hot gas components.

\underbar{NGC~3818} ~~~ Scorza et al. (\cite{Scorza98}) surface photometry
suggests that this galaxy
belongs to the class of bulge-dominated early-types. In particular, NGC~3818 
is representative of the class, 
having a disk fully embedded in a boxy bulge. The 
disk profile is exponential all throughout the object.  Adopting the values of 
$V_M$=114$\pm$7 km~s$^{-1}$ and $\sigma$=199$\pm$10 given by Simien \&
Prugniel (\cite{SP97}) in order to obtain a rough estimate of the 
position in the (V$_m$/$\sigma$, $\epsilon$) plane, we may consider  the galaxy 
not far from the region of the oblate galaxies with (nearly) 
isotropic velocity dispersion.
The bulge boxiness, anyway, suggests a relative peculiarity of this galaxy 
as indicated by the $\Sigma$=1.30 assigned 
by Sansom et al. (\cite{Sansom00}).   

\underbar{NGC~4374, M~84}~~~ Caon et al. (\cite{Caon90}) provide the
B-band luminosity profile of this galaxy extending up $\approx$12
r$_e$ (at 26 $\mu_B$=26 mag~arcsec$^{-2}$ the galaxy outskirts overlap
with those of NGC~4406 (M~86)). The isophote has  a nearly constant
position angle up to $\approx$80\arcsec (1.6 r$_e$) and a strong
variation ($\Delta$ P.A.$\approx$50$^\circ$) in the outskirts where
there are possible contamination by NGC~4406. 
They notice a complex dust-lane system in the inner 5\arcsec\ which
is beautifully visible in the WFPC2 images (see e.g. Bower et al.
\cite{Bower00}). 

The galaxy stars rotate very slowly ($\leq$ 38 km~s$^{-1}$ (Davies et al.
\cite{Davies83} and other rotation velocity determinations in {\tt
HYPERCAT}) which leads to a very low V$_m$/$\sigma$$\leq$0.15, indicating
the presence of anisotropy in the velocity dispersion.

WFPC2 high resolution images of NGC~4374 confirm the presence of an 
extended central ionized gas component (Bower et al. \cite{Bower00} and
reference therein). In the inner 5$\arcsec$ the emission, detected in
H$\alpha$+[NII], has three components: a nuclear disk, a ionization
cone and outer filaments. The ionization cone is similar to those found 
in Seyfert galaxies and is also aligned with the radio axes. NGC~4374
is indeed also a Fanaroff-Riley  type I (FRI) radio source (Laing \&
Bridle \cite{Laing87}). 

NGC~4374 is then one of the nearby BH candidate galaxies with a
relatively week AGN. The kinematics of the nuclear disk of ionized gas
indicates the presence of a central 1.5$\times$10$^9$ M$_\odot$ dark
compact object. 

Finoguenov \& Jones (\cite{Fino01}) analyzed deep {\it Chandra}
observations of NGC~4374 finding a central AGN, several galactic
sources and a diffuse hard emission, where the gas is probably heated
by the central AGN. The soft emission instead has the same spatial
distribution of the   radio structure of the galaxy. 
 
\underbar{NGC~4696}~~~ This galaxy, the dominant member of the 
Centaurus Cluster, is know to be a radio galaxy PKS1246-41, with a  FRI type. 
Together with the ionized gas, with line ratios typical of a LINER (see Lewis
et al. \cite{Lewis03}), it possesses neutral gas. This latter is kinematically
associated with the compact emission
filament system and the dust-lane present in the central 20\arcsec\ 
of the galaxy (Sparks et al.\cite{Sparks97}). 
Allen at al. (\cite{Allen00}) detected  hard X-ray emission
components in the spectra of NGC~4696. The characteristics of 
the emission are different from  
those of Seyfert galaxies, the latter having steeper power-law components and
higher X-ray luminosities.  
They argue that the hard X-ray emission is likely to be caused 
by accretion onto a central, supermassive BH. 

The large scale environments of this galaxy is perturbed: ASCA observations 
show evidence that the main cluster, centred on NGC~4696, is strongly 
interacting/merging with a sub-cluster centered on NGC~4709 
(see Churazov et al. \cite{Churazov99}).    

\underbar{NGC~4697}~~~ The galaxy is located in the Virgo Cluster southern
extension (Tully \cite{Tu88}). The surface photometry of Scorza et al.
(\cite{Scorza98}) shows that the galaxy belongs to the disky family of Es,
the faint disk being visible throughout the luminosity profile. In
agreement with this finding, no rotation has been measured along the
minor axis by Bertola et al. (\cite{Bertola88}). Goudfrooij et al.
(\cite{Gou94}) reported a significant detection in both H$\alpha$ and
[NII] lines with an extension of $\approx$35\arcsec.

NGC~4697 is considered a X-ray faint early-type galaxy. 
Sarazin et al. (\cite{Sarazin01}) observed the galaxy with 
{\it Chandra}, resolving much of the X-ray emission (61\%) into 90
point sources, mostly LMXBs which have lost much of their interstellar gas. 
The galaxy center hosts a X-ray source which could be an active galactic 
 nucleus (but also one or more LMXBs)
 radiating at a very small fraction ($\leq$ 4 $\times$ 10$^{-8}$) of its 
 Eddington luminosity.   

\underbar{NGC~5044}~~~ The galaxy, located in a rich group of galaxies
(see e.g. Tully \cite{Tu88}), is rich of dust in the central 10\arcsec\ with
a clumpy distribution, NGC~5044 has been detected by IRAS and by ISO
(Ferrari et al. \cite{Ferrari02} and reference therein). The gas has a filamentary shape
with an extension of about 40\arcsec\ (Macchetto et al. \cite{Mac96}).
The gas velocity profile is irregular, with many humps and dips while
the inner (within 1/3 of the effective radius) stellar velocity profile
is conter-rotating with respect to the outer regions (Caon et al.
\cite{CMP00}). The galaxy then configures as a very peculiar
object being a possible merger/accretion remnant. 
 
Recently, Rickes et al. (\cite{Rickes04}) studied the ionized gas
component in this galaxy suggesting the presence of both a non-thermal
ionization source in the central region and an additional ionization
source (possibly hot post-AGB stars) in the outer parts.

\underbar{NGC~5090}~~~ The galaxy together with NGC~5091 form the pair
RR~242 (Reduzzi \& Rampazzo \cite{ReRa95}) which is part 
of a loose group. Considering a radius of about half a degree
around NGC~4105, there are four additional luminous galaxies
with comparable redshift: NGC~5082 (separation 5.8\arcmin\ 77.5 kpc,
$\Delta$ V$_{pair}$= 421 km~s$^{-1}$), NGC~5090A (separation 20.3\arcmin\
298.4 kpc, $\Delta$ V$_{pair}$= 45 km~s$^{-1}$), NGC~5090B (separation
13.8\arcmin\ 184.1 kpc, $\Delta$ V$_{pair}$= 773 km~s$^{-1}$) and
ESO~270~G007 (separation 24.4\arcmin\ 326.1 kpc, $\Delta$ V$_{pair}$= 275
km~s$^{-1}$). The group could probably include also ESO~270~G003
(separation 3.5\arcmin\ 46.7 kpc) whose  redshift is still unknown.
Considering the above galaxies, the possible loose group has
an average recession velocity of  3681 km~s$^{-1}$ and a
velocity dispersion of 327 km~s$^{-1}$.
NGC~5090 seems a``bona fide'' elliptical without
particular signatures of interaction, according to the surface photometry
of  Govoni et al. (\cite{Govo00}). 
The galaxy hosts an FRI radio source (PKS B1318-434) 
with two large radio jets and the radio axis perpendicular to the
line connecting the nuclei of the pair members (see Llyod et al. \cite{Lloyd96}
and reference threin). Carollo et al. (\cite{CDB93}) have obtained 
the velocity dispersion and rotation velocity profiles of NGC~5090
showing a high central velocity dispersion and a low rotation
velocity both characteristic of E galaxies. Bettoni et al. 
(\cite{Bettoni03}), in their
study of the BH mass of low redshift radio galaxies,
attributed to NGC~5090 a BH mass of 1.1$\times$10$^9$ M$_\odot$.

\underbar{NGC~5193}~~~ The galaxy forms a physical pair with
NGC~5193A, two S0s according to the ESO-LV classification.   
Reduzzi \& Rampazzo (\cite{ReRa95}) showed that NGC~5193 is
an E, with disky isophotes, while the companion is probably and S0a
showing incipient arms. The ongoing interaction of the two galaxies
is shown by the warped disk of NGC~5193A. The small disk in NGC~5193
seems aligned with that of the companion, as shown by Fa\'undez-Abans
\& de Oliveira-Abans (\cite{FdeO98}).  

\underbar{NGC~5266}~~~ The galaxy has a prominent dust-lane along its
projected minor axis. The kinematics study of Varnas et al. (\cite{Va87})
shows that stars rotate about the optical minor axis 
while the gas in the dust-lane rotates about the optical major axis, 
i.e. the kinematic axes of the
stars and gas appear othogonal. Varnas et al. (\cite{Va87}) suggest
that the underlying galaxy is triaxial. 

The ionized gas in NGC~5266 lies in a ring associated with the dust ring
Goudfrooij (\cite{Gou94}). Sage \& Galletta
showed (\cite{SG93}) that the CO has
a ring like distribution and that the molecular and ionized gas are
co-rotating. Morganti et al. (\cite{Morganti97}) detected
neutral gas  up to 8 times the optical effective
radius each side of the galaxy. The outer HI gas extends almost
orthogonal to the optical dust--lane: the overall HI kinematics can be
modeled by assuming that the gas lies in two orthogonal planes, the
plane of the dust--lane, in the central parts, and that perpendicular to
this in the outskirts. The large amount of gas, more than 10$^{10}$
M$_\odot$, and the HI morphology suggest that the galaxy could
be the remnant of an old merging episode between two spiral galaxies,
since the HI gas appears settled.
 
\underbar{NGC~5638}~~~ The galaxy forms a physical pair with NGC~5636, 
an SBa at 2.3\arcmin\ 
separation. The systemic velocity difference between the two galaxies is
$\Delta$V=69 km~s$^{-1}$. According to the surface photometry of
Peletier et al. (\cite{Pele90}), NGC~5638  is a rare elliptical with
``truly'' elliptical isophotes since no significant 
boxy or disk-like deviations are visible in the $a_4$ 
shape profile. The position angle variation is $\leq$30$^\circ$ while
the (B-R) color is quite stable in the range $\approx$ 1.6--1.5 along
all the galaxy. The galaxy kinematics along the major axis has 
been studied by Davies et al.
(\cite{Davies83}). They found that at about 15\arcsec\ from the nucleus
the galaxy projected rotation velocity is   $\approx$90 km~s$^{-1}$ with
a velocity dispersion of $\approx$120 km~s$^{-1}$. 
The position of the galaxy in the (V$_m$/$\sigma$, $\epsilon$) plane is 
consistent with the line traced by oblate galaxies with isotropic
velocity dispersion. According to its photometric and kinematical
properties, this galaxy should represent a 
typical ``normal'' elliptical galaxy.

\underbar{NGC~5812}~~~ Recently, high resolution images have been obtained
with WFPC2 (Rest et al. \cite{Rest01}). 
NGC~5812 has a small dust disk which extends for 0.4\arcsec.
The luminosity profile has a "power law" shape.

NGC~5812 shows a declining stellar velocity dispersion profile
and little rotation ($\leq$ 40 km~s$^{-1}$ up to a radius of
25\arcsec, i.e. at about 1 r$_e$ (Bertin et al. \cite{B94}).

\underbar{NGC~5813}~~~~ Caon et al. (\cite{CMP00}) kinematic study
confirms the presence of a decoupled stellar core. The ionized gas 
has an irregular velocity profile along both the studied directions,
suggesting that the gas is still unsettled, a conclusion which
is also supported by the filamentary morphology of H$\alpha$+[NII]
emission (see also \cite{Mac96}). The WFPC2 image of NGC~5813
suggests the presence of dust in the inner 10\arcsec\ and a "core-type"
luminosity profile (Rest et al. \cite{Rest01}).

\underbar{NGC~5831}~~~ Reid et al. (\cite{Reid94}) multicolor photometry
indicates that the luminosity profile follows in an excellent way the
r$^{1/4}$ law at all radii and that the galaxy does not show
statistically significant evidence for monotonic changes in color with
radius. High resolution images have been obtained with WFPC2. 
NGC~5831 does not have dust-lane and the luminosity profile has  a
"power law" shape (Rest et al. \cite{Rest01}) and a significant twisting
($\approx$20$^\circ$) in the inner 20\arcsec.

Davies et al. (\cite{Davies83}) show that stars in NGC~5831 
rotate very slowly along major axis (27$\pm$9 km~s$^{-1}$): 
the V/$sigma$ value is 0.18 indicating the presence
of anisotropies in the velocity dispersion. 

\underbar{NGC~6758} ~~~ Caon et al. (\cite{CMP00}) 
measured the gas and stars velocity profiles along P.A.=63$^\circ$
an P.A.=153$^\circ$. Along P.A.=63$^\circ$ the gas velocity profile is  regular but,
after reaching a maximum of about 140 km~s$^{-1}$, decreases up to V=0.
The stars rotate slowly (about 40 km~s$^{-1}$) and in the opposite sense
with respect to the gas. Along P.A.=153$^\circ$ gas and stars rotate in the
same sense. The gas could be then accreted from the outside.

\underbar{NGC~6776}~~~The galaxy shows a diffuse,
asymmetric luminosity distribution and  tails.
Malin \& Carter (\cite{MC83})  annotated in their catalogue that the fine
structure characterizing  NGC~6776 seems different from classical shells
and it is more reminiscent of a tidal debris forming a loop and tails.
The surface photometry made by Pierfederici \& Rampazzo (\cite{PR04}) 
suggests an E/S0 classification. The isophotes appear strongly
twisted in the central region ($<$ 20\arcsec)  and then stabilize around
P.A.=20$^{\circ}$, suggesting either the presence of a triaxial structure
or the presence of two distinct structures.  The shape of the profile of
the a$_4$ parameter indicates the presence of  boxy isophotes, but could
also be influenced by the presence of the numerous foreground stars
(although they have been properly masked) in addition to intrinsic
asymmetries/tails of this galaxy.  Pierfederici \& Rampazzo (\cite{PR04}) 
show that the prominent tail, visible in the E side of the galaxy and
extending in the NS direction, has color 
(V-R)=0.62$\pm$0.03 consistent with the average color of the galaxy
$\langle$V-R$\rangle$=0.58$\pm$0.04.

Macchetto et al. (\cite{Mac96}) found that the ionized gas component in
NGC 6776 has  a filamentary structure suggesting that the gas has
not yet settled in the galaxy potential. The inner (10\arcsec\ wide)
kinematics of NGC~6776 (Longhetti et al. \cite{L98b}), shows a
regular rotation curve, while the velocity dispersion profile in the
same region is asymmetric and shows a plateau ($\approx$ 250 km~s$^{-1}$)
on the NE side.   The study of line strength indices in Longhetti et al.
\cite{L99,L00} suggests that the galaxy had a recent burst ($<$10$^9$ years)
of  star formation.

\end{document}